\documentclass[twocolumn,showpacs,graphicx,floatfix,eqsecnum,prb,twocolumn,hyperref=pdftex,amsmath,longbibliography]{revtex4-1}
\usepackage{bm}
\usepackage{enumerate}
\usepackage{graphicx}
\usepackage[dvips]{epsfig}
\usepackage{epsf}
\usepackage{epstopdf}
\usepackage{hyperref}
\usepackage{comment}
\hypersetup{colorlinks=true,linkcolor=blue,anchorcolor=blue,citecolor=blue,filecolor=blue,urlcolor=blue,bookmarksnumbered=true,pdfview=FitB}
\usepackage{fancyhdr}
 
\newcommand{\Rmnum}[1]{\expandafter\@slowromancap\romannumeral #1@}
\newcommand{\beq}{\begin{equation}}
\newcommand{\eeq}{\end{equation}}
\newcommand{\bea}{\begin{eqnarray}}
\newcommand{\eea}{\end{eqnarray}}
\newcommand{\e}{\varepsilon}

\newcommand{\vf}{v_{\mathrm{F}}}
\newcommand{\kf}{k_{\text{F}} }
\newcommand{\Ef}{E_{\text{F}} }

\newcommand{\bk}{{\bf k}}
\newcommand{\bp}{{\bf p}}
\newcommand{\bq}{{\bf q}}
\newcommand{\bQ}{{\bf Q}}
\newcommand{\bu}{{\bf u}}
\newcommand{\bv}{{\bf v}}

\newcommand{\nn}{\nonumber}
\newcommand{\bwt}{\begin{widetext}}
\newcommand{\ewt}{\end{widetext}}
\newcommand{\im}{\mathrm{Im}}
\newcommand{\bse}{\begin{subequations}}
\newcommand{\ese}{\begin{subequations}}
\newcommand{\re}{\text{Re}}
\newcommand{\bs}{\boldsymbol{\sigma}}
\newcommand{\br}{{\bf r}}

\begin{document}
\title{Dynamical 
susceptibility of a Fermi liquid}
\author{Vladimir A. Zyuzin$^{a,b}$, Prachi Sharma$^a$, and Dmitrii L. Maslov$^a$}
\affiliation{$^a$Department of Physics, University of Florida, P.O. Box 118440, Gainesville, Florida 32611-8440, USA\\
$^b$Department of Physics and Astronomy, Texas A$\&$M University, College Station, Texas 77843-4242, USA}
\begin{abstract}
We study dynamic response of a Fermi liquid in the spin, charge and nematic channels beyond the random phase approximation for the dynamically screened Coulomb potential. In all the channels, one-loop order corrections to the irreducible susceptibility result in a non-zero spectral weight of the corresponding fluctuations above the particle-hole continuum boundary. It is shown that the imaginary part of the spin susceptibility,  $\im\chi_{s}(\bq,\omega)$, falls off as $q^2/\omega$ for frequencies above the continuum boundary ($\omega\gg \vf q$) and below the model-dependent cutoff frequency, whereas the imaginary part of the charge susceptibility, $\im\chi_c(\bq,\omega)$, falls off as $(q/k_F)^2 q^2/\omega$ for frequencies above the plasma frequency.  An extra factor of $(q/k_F)^2$ in $\im\chi_c(\bq,\omega)$ as compared to $\im\chi_{s}(\bq,\omega)$ is a direct consequence of Galilean invariance. The imaginary part of the nematic susceptibility increases linearly with $\omega$ up to a peak at the ultraviolet energy scale-- the plasma frequency and/or Fermi energy--and then decreases with $\omega$. We also obtain explicit forms of the spin susceptibility from the kinetic equation in the collisionless limit and for the Landau function that contains up to first three harmonics.

\end{abstract}
\date{\today}
\maketitle

\section{Introduction}
\label{sec:intro}
The dynamical susceptibility $\chi(\bq,\omega)$ of an interacting Fermi system, and, in particular, its imaginary part, is a fundamental quantity which contains the information about  the strength of fluctuations of a particular order parameter. The charge ($c$) and spin ($s$) dynamical susceptibilities can be measured directly by a number of experimental techniques, such as inelastic electron \cite{mitrano:2017} and neutron \cite{lovesey} scattering, Raman scattering (both in parallel and cross-polarization geometries),\cite{devereaux:2007} and inelastic  X-ray scattering. \cite{ament:2011} In addition, $\chi_{c,s}(\bq,\omega)$ determines the dispersions and damping of the collective modes, such as zero-sound and plasmon modes in the charge channel, and the Silin-Leggett mode\cite{silin:1958,leggett:1970} of a partially spin-polarized Fermi liquid (FL) or the magnon mode\cite{white,ma:1968}  of a ferromagnetic FL in the spin channel. Exchange by ferromagnetic fluctuations, whose spectrum is parameterized by $\chi_s(\bq,\omega)$, is believed to be the main pairing mechanism in superfluid $^3$He and ferromagnetic superconductors. Finally, interaction of itinerant fermions with critical magnetic fluctuations is responsible for the breakdown of FL near a ferromagnetic quantum phase transition.

Dynamical response of a FL is well-understood in the hydrodynamic limit,\cite{nozieres} i.e., at frequencies satisfying the condition $\omega\tau_{\text{qp}}\ll 1$, where $\tau_{\text{qp}}$ is the quasiparticle scattering time at finite $T$. This regime can be described with a minimal knowledge about the mutual scattering of quasiparticles: all is required from this scattering is to be frequent enough to establish local equilibrium.
However, many experiments and, in particular the most recent inelastic electron study of a copper-oxide superconductor,\cite{mitrano:2017}are performed at high enough frequencies and low enough temperatures so that the system is in the collisionless regime, which corresponds to  $\omega\tau_{\text{qp}}\gg 1$.

There are much fewer theoretical results for dynamical response of a FL  in the collisionless regime. Typically, dynamical susceptibilities  in this regime are calculated within the random phase approximation (RPA), which amounts to resumming the chains of 
free-fermion polarization bubbles in the charge channel or the ladder diagrams in the spin channel, or by solving the FL kinetic equation without the collision-integral term.
There are two well-known results in this limit: \cite{nozieres,physkin,baym:book} one is that the imaginary part of either charge or spin susceptibilities
scales as $\omega/q$ for $\omega\ll v_Fq$, i.e., well below the boundary of the particle-hole continuum, and another one is that the real part of the susceptibility scales as $q^2/\omega^2$
for $\omega\ll v_Fq$, i.e., well-above the continuum boundary.  [In the charge channel, by "susceptibility" we understand its irreducible part.] These scaling forms are the same as for free fermions except for the prefactors which depend on the Landau parameters of a FL.
In this approximation, which completely neglects the residual interaction between quasiparticles or, equivalently, considers an excitation of  a single-particle hole pair, the imaginary part of any susceptibility is strictly zero outside the particle-hole continuum.

For many purposes, however, one is interested in the spectral weight of particle-hole excitations, $\im\chi_{c,s}(\bq,\omega)$, outside the continuum.  To get a non-zero
$\im\chi_{c,s}(\bq,\omega)$ in this region, 
one needs to take into account the residual interaction between quasiparticles or, equivalently, excitation of multiple particle-hole pairs. Such processes in the charge channel were analyzed in the context of plasmon attenuation outside the particle-hole continuum \cite{dubois:1969,mishchenko:2004} and renormalization of the dielectric function of graphene\cite{sodemann:2012} but, to the best of our knowledge, the spin channel has not been considered in the prior literature. The nematic susceptibility, i.e., a susceptibility of a non-conserved order parameter, has recently been considered outside the continuum in Ref.~\onlinecite{klein:2018}.

\begin{figure}[t]
\includegraphics[width=1.0 \linewidth]{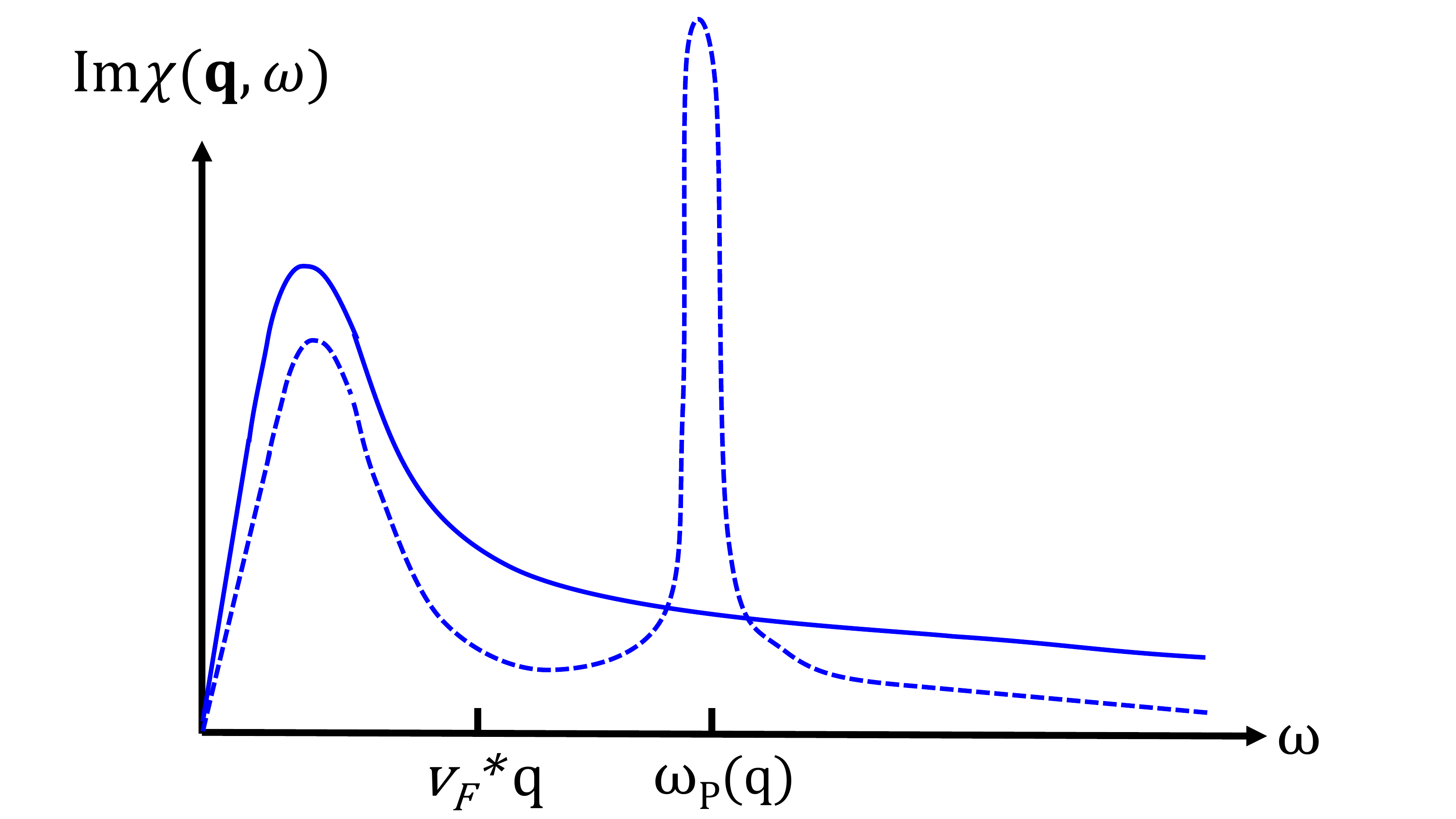}
\caption{A sketch of the imaginary part of the spin (solid) and charge (dashed) susceptibilities.
The Bohr magneton is set to zero, such that the units of $\chi_s$ and $\chi_c$ are the same. 
$\omega_p(q)$ denotes the plasmon frequency which may depend on $q$.  The particle-hole continuum occupies the range $0\leq \omega\leq \vf^* q$. Outside the continuum, $\im\chi_s(\bq,\omega)$ falls off a $q^2/\omega$ both in 2D and 3D.  For $\omega\gg\omega_p(q)$,  the tail of $\im\chi_c(\bq,\omega)$ is smaller than that of $\im\chi_s(\bq,\omega)$ by a factor of $(q/\kf)^2\ll 1$, which reflects Galilean invariance of the system. In the intermediate range, $\vf^*q\ll \omega\ll\omega_p(q)$, $\im\chi_c(\bq,\omega)$  increases with $\omega$ as $q^2\omega^3$ and $q^4\omega^3$ in 2D and 3D, respectively. The relative
magnitude of $\chi_s$ and $\chi_c$ within the continuum depends on the interaction, and the choice made in the sketch is completely arbitrary.\label{fig:spin-charge}}
\end{figure}

In this paper, we derive a number of explicit results concerning dynamical response of a FL, at the level of both non-interacting and interacting quasiparticles. The latter case is considered at one-loop order in the dynamically screened Coulomb interaction.
We assume that $T=0$ and that disorder is negligible, i.e., that $\min\{\omega,\vf q\}\gg \max\{T,1/\tau_{\text{d}}\}$, where $\tau_{\text{d}}$ is the mean free time due to disorder scattering.
With these assumptions, our main results are as follows:
\begin{itemize}
\item[a)] for frequencies well above the continuum boundary but below some model-dependent cutoff, i.e., for $\vf q\ll \omega\ll \omega_0$, the imaginary part of the spin susceptibility falls off in a universal manner: $\im\chi_{s}(\bq,\omega)\propto q^2/\omega$ both for $D=2$ and $D=3$. For $\omega\gg \omega_0$, $\im\chi_{s}(\bq,\omega)$ falls off faster than $1/\omega$.
A sketch of $\im\chi_{s}(\bq,\omega)$ as a function of $\omega$ is shown by the solid line in Fig.~\ref{fig:spin-charge};

\item[b)] in a Galilean-invariant system and for frequencies well above the plasmon mode, $\omega_p(q)$, the imaginary part of the charge susceptibility is suppressed by a factor of $(q/\kf)^2\ll 1$ as compared to the spin one, i.e., $\im\chi_{c}(\bq,\omega)\propto (q/\kf)^2q^2/\omega$. 
[A sketch of $\im\chi_{c}(\bq,\omega)$ as a function of $\omega$ is shown by the dashed line in Fig.~\ref{fig:spin-charge}.]
An extra factor of $(q/\kf)^2\ll 1$ reflects the fact that the real part of the conductivity of a Galilean-invariant FL must vanish at $q=0$. \cite{mishchenko:2004} On a technical level, the relative suppression of the charge susceptibility compared to the spin one occurs as a result of partial cancellation between the self-energy, ladder, and Aslamazov-Larkin (AL) diagrams. Namely, the $q^2/\omega$ term in $\im\chi_{s}(\bq,\omega)$, which comes from the sum of the self-energy and ladder diagrams,  is canceled by the same term from the AL diagrams, which contribute    to $\chi_c$ but not to $\chi_s$. It is interesting to note that the same suppression of charge fluctuations relative to spin ones occurs also in 1D. \cite{zyuzin:2014}

 \item[c)] In the intermediate range of frequencies, $\vf q\ll\omega\ll \omega_p(q)$, $\im\chi_{c}(\bq,\omega)$ raises towards a plasmon peak at $\omega=\omega_p(q)$ as $q^2\omega^3$ in 2D and as $q^4\omega^3$ in 3D. 
 \end{itemize}
 
 The $q^2/\omega$ asymptotic form of $\im\chi_{s}(\bq,\omega)$ 
 outside the continuum can be obtained by the following simple argument. Spin conservation and analyticity require that  $\chi_s(\bq,\omega)\propto q^2$ at $q\to 0$; hence the factor of $q^2$ follows immediately. The $1/\omega$ dependence is the first non-vanishing term in the high-frequency expansion, which is consistent with the requirement that $\im\chi_s$ must be an odd function of $\omega$. In 2D, the combination $q^2/\omega$ already has the units of the density of states; hence there is no room for more dimensional parameters, and the final result is given by $q^2/\omega$ multiplied by a dimensionless coupling constant. In 3D, one needs an additional factor with the units of momentum, which is provided by $k_{\text{F}} $. 

Nevertheless, the same argument does not work for the charge susceptibility of  a Galilean-invariant system, whose tale behaves as $q^4/\omega$ rather than $q^2/\omega$. As we said above, an extra factor of $q^2$ comes from the requirement that the optical conductivity of a Galilean-invariant FL by itself vanishes as $q^2$ (Ref.~\onlinecite{mishchenko:2004}).

The rest of the paper is organized as follows. 
In Sec.~\ref{sec:rpa}, we analyze dynamic susceptibilities of a Fermi liquid at the level of non-interacting quasiparticles, i.e., within the kinetic equation with zero right-hand side. In Sec.~\ref{sec:any}, we go beyond the level of non-interacting quasiparticles and calculate the spin (Sec.~\ref{sec:spin}), charge (Sec.~\ref{sec:charge}), and nematic (Sec.~\ref{sec:nematic}) susceptibilities to one-loop order in the dynamically screened Coulomb interaction. Our conclusions are given in Sec.~\ref{sec:concl}. Some technical details of the calculations are delegated to Appendix \ref{sec:appb}.

\section{
Dynamic susceptibility\\ of a Fermi liquid:\\
non-interacting quasiparticles}
\label{rpa}
\subsection{Random Phase Approximation}
\label{sec:rpa1}
For completeness, we remind the reader of well-known RPA results for the charge and spin susceptibilities.
 For electrons interacting via the Coulomb potential $U_0(\bq)=2\pi e^2/q$ (in 2D) and $U_0(\bq)=4\pi e^2/q^2$ (in 3D), the RPA form of the charge susceptibility is \cite{mahan:book}
\bea
\chi_c(\bq,\omega)=\frac{\chi^{(0)}(\bq,\omega)}{1+U_0(\bq)\chi^{(0)}(\bq,\omega)},
\label{chi_charge}
\eea 
where $\chi^{(0)}(\bq,\omega)$ is the free-electron susceptibility.
The spin susceptibility, obtained by resumming the ladder series for a Hubbard-like interaction with coupling constant $U$, is given by \cite{white}
\beq
\chi_{s}(\bq,\omega)=\frac{\chi^{(0)}(\bq,\omega)}{1-\frac{U}{2}\chi^{(0)}(\bq,\omega)}.\label{hubbard}
\eeq

\begin{figure}[t]
\includegraphics[width=0.7 \linewidth, angle = -90]{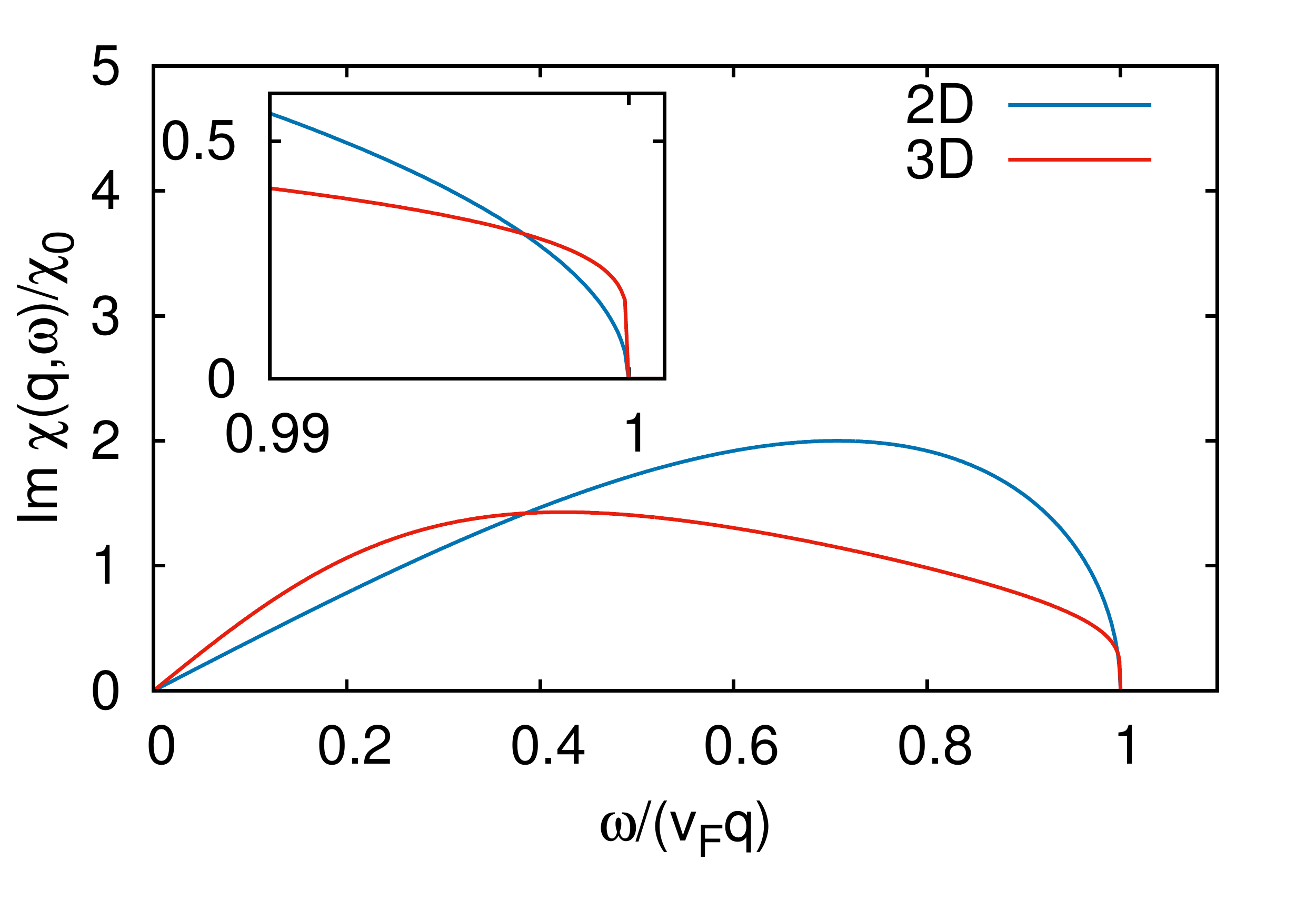}
\caption{Imaginary part of the spin susceptibility, $\chi_s(\bq,\omega)$, within the RPA [Eq.~(\ref{hubbard})].  The inset shows $\im\chi_s(\bq,\omega)$ near the boundary of the particle-hole continuum ($\omega = \vf q$), where it vanishes as $\sqrt{\vf q-\omega}$ and  $1/\ln^2(\vf q-\omega)$ in 2D and 3D, respectively. The dimensionless coupling constant is $UN_F/2 = 0.5$ for both dimensions, where $N_F$ is the density of states at the Fermi level. \label{fig:rpa}}
\end{figure}

 The imaginary part of $\chi^{(0)}(\bq,\omega)$ is non-zero only within the particle-hole continuum, i.e., for $\omega<\vf q$ (assuming that $q\ll \kf$). Within the RPA, the same is also true for $\im\chi_{c,s}(\bq,\omega)$. The vanishing of $\im\chi_{c,s}(\bq,\omega)$ at the continuum boundary
as $\sqrt{\vf q-\omega}$ in 2D and as $1/\ln^2(\vf q-\omega)$ in 3D, reflects the corresponding threshold singularities of $\chi^{(0)}(\bq,\omega)$. A profile of $\im\chi_s(\bq,\omega)$ within the RPA is shown in Fig.~\ref{fig:rpa}.
The imaginary part of the charge susceptibility 
is qualitatively similar to that shown in Fig.~\ref{fig:rpa}, except for a sharp peak above the continuum, which corresponds
either to zero sound mode (for neutral fermions) or to a plasmon (for electrons). 

\subsection{Collisionless kinetic equation for a Fermi liquid}
\label{sec:FLspin}

In this section, we analyze the charge and spin susceptibilities of a FL 
at the level of non-interacting quasiparticles. 
Technically, this amounts to solving the kinetic equation in the presence of a time- and position-dependent external perturbation but without the collision integral.\cite{nozieres,physkin,baym:book} 
This approach is identical to summing up the ladder series, in which polarization bubbles formed by quasiparticles are separated by irreducible interaction vertices. \cite{eliashberg:1962,finkelstein:2010,chubukov:2014,wu:2018} We will find analytic solutions of the kinetic equation
for several model forms of the Landau function with a finite number of harmonics and solve the kinetic equation numerically for the case of Coulomb interaction. The main goal of this section is to obtain
more accurate results for the spin susceptibility than the one given by RPA  with the Hubbard-like interaction, Eq.~(\ref{hubbard}), which corresponds to an isotropic Landau function in the FL theoy. 

Since the case of a perturbation in the charge channel is analyzed in Refs.~\onlinecite{nozieres} and \onlinecite{baym:book}, we discuss the spin channel in detail and give only the results for the charge channel later. The starting point for the spin channel is the collisionless kinetic equation for a  FL in the presence of a weak, time-dependent and non-uniform magnetic field, ${\bf B}(\bf r,t)$:
\beq
\left(\partial_t+v^*_F\hat\bk\cdot\boldsymbol
{\nabla}_{\bf r}\right)\delta \hat n_\bk-\vf^*\hat\bk\cdot\boldsymbol{\nabla}_{\bf r}\delta\hat\epsilon_\bk n'_0=0,
\label{ke}
\eeq
where $\vf^*$ is the renormalized Fermi velocity, $n_0(\e)$ is the equilibirum Fermi function, $n'_0\equiv\partial_\e n_0(\e)$, $\hat n_{\bk}$ is the occupation number, and  $\hat\epsilon_\bk$ is the quasiparticle energy (the last two quantities being $2\times 2$ matrices). The commutator $i\left[\hat\epsilon_\bk,\hat n_\bk\right]$, which describes precession of electron spins around the direction of the external magnetic field,\cite{lifshitz:1980} can be neglected in the linear-response regime.
As usual, a variation of the quasiparticle energy is decomposed into two parts: one is the Zeeman splitting due to the magnetic field (for a charged FL, the magnetic field is assumed not to affect the orbital motion of quasiparticles) and another one is due to a self-consistent field from other quasiparticles
\beq
\delta\hat\epsilon_\bk(\br,t)=-{\bf B}(\br,t)\cdot\bs+\text{Tr}'\int\frac{d^Dk'}{(2\pi)^2}{\hat f}(\bk,\bk')\delta \hat n_{\bk'}(\br,t),
\label{deltae}
\eeq
where ${\hat f}(\bk,\bk')=\hat I\hat I'f^s(\hat\bk\cdot\hat\bk')+\bs\cdot\bs' f^a(\hat\bk\cdot\hat\bk')$ is the Landau function, and Tr$'$ goes over the spin variables of the ``primed" quasiparticle. The non-equilibrium part of the distribution function can be expanded over a complete set of Pauli matrices
\beq
\delta\hat n_{\bk}(\br,t)=-n_0'{\bf u}(\hat\bk;\br,t)\cdot\bs.
\label{deltan}
\eeq
Substituting Eq.~(\ref{deltae}) and (\ref{deltan}) into Eq.~(\ref{ke}), evaluating the traces, and switching to the momentum-space representation, we obtain
an integral equation for ${\bf u}(\hat\bk;\bq,\omega)$ (the last two arguments in ${\bf u}$ will be omitted for brevity)
\bea
{\bf u}(\hat\bk)&=&{\cal P}(\hat k;\bq,\omega)\left[-{\bf B}(\bq,\omega)
+\int \frac{d\hat\bk'}{{\cal O}_D} F^a(\hat\bk\cdot\hat\bk'){\bf u}(\hat\bk')\right],\nn\\
\label{ueq}
\eea
where $F^a(\hat\bk\cdot\hat\bk')=N_F^*f^a(\hat\bk\cdot\hat\bk')$,  $N_F^*$ is the renormalized density of states, ${\cal O}_D$ is the full solid angle in $D$ dimensions, and
\beq
\label{Pi}
{\cal P}(\hat\bk;\bq,\omega)=\frac{\vf^*\hat\bk\cdot\bq}{\omega-\vf^*\hat\bk\cdot\bq+i0^+}
\eeq 
is a propagator of a particle-hole pair with the total momentum $\bq$ formed by a particle and hole, moving in the directions of $\hat\bk$ and $-\hat\bk$, respectively. 
 Since ${\cal P}(\hat k;\bq,\omega)$ depends on the angle $\theta_{\bk}$ between $\hat\bk$ and $\bq$,  it can be expanded over the complete set of angular harmonics
\bea
{\cal P}(\hat k;\bq,\omega)=\left\{\begin{array}{ccc}\sum^\infty_{\ell=-\infty} e^{i\ell\theta_\bk}{\cal P}_\ell(q,\omega),\\
\\
\sum_{\ell=0}^\infty (2\ell+1) P_\ell(\cos\theta_\bk){\cal P}_\ell(q,\omega),
\end{array}
\right.
\label{Pexp}
\eea
in 2D and 3D, respectively, with $P_\ell(x)$ being the Legendre polynomial. 

In 2D, the harmonics ${\cal P}_{\ell}(q,\omega)$  are given by\cite{wu:2018}
\begin{subequations}
\bea
{\cal P}_\ell(q,\omega)&=&-\delta_{\ell,0}+z\int^\pi_0 \frac{d\theta_\bk}{\pi}\frac{\cos(\ell\theta_\bk)}{z-\cos\theta_\bk+i 0^+}
\label{Pl2D}\\
&=&-\delta_{\ell 0}+\left\{
\begin{array}{ccc}
 (-i)^{|\ell|+1}e^{i|\ell|\psi}\frac{z}{\sqrt{1-z^2}},\,z<1;\\
 e^{-|\ell| \psi} \frac{z}{\sqrt{z^2-1}},\; z>1,
 \end{array} 
\right.
\label{Pl2Dexp}
\eea
\end{subequations}
where 
\bea
z\equiv \omega/\vf^*q,
\label{z_def}
\eea
 $\sin\psi=z$ for $z<1$ and $\sinh\psi =z$ for $z>1$. Without loss of generality, we take $z$ to be non-negative.
In what follows, we will need explicit forms of the few first harmonics within the continuum ($z<1$):
\begin{subequations}
\bea
{\cal P}_0(q,\omega)&=&-1-i\frac{z}{\sqrt{1-z^2}},\label{P02}\\
{\cal P}_1(q,\omega)&=&-z\left(1+i\frac{z}{\sqrt{1-z^2}}\right),\label{P12}\\
{\cal P}_2(q,\omega)&=&-2z^2+i\frac{z}{\sqrt{1-z^2}}\left(1-2z^2\right).\label{P22}
\eea
\end{subequations}
In 3D, the harmonics ${\cal P}_{\ell}(q,\omega)$ are given by \cite{gradshteyn,baym:book}
\begin{subequations}
\bea
{\cal P}_\ell(q,\omega)&=&-\delta_{\ell,0}
+z\int^1_{-1} \frac{dy}{2}\frac{P_\ell(y)}{z-y+i 0^+}\label{Pl3D}\\
&=&-\delta_{\ell 0}+
z\left[Q_\ell(z)-i\frac{\pi}{2} P_\ell(z)\theta(1-z)\right],\nn\\
\label{Pl3Dexp}
\eea
\end{subequations}
where $Q_\ell(x)$ for $|x|<1$ is the Legendre function of the second kind, i.e., the second linearly independent solution of the Legendre differential equation. For $|x|>1$,  $Q_\ell(x)$ is to be understand as an analytic continuation of the Legendre function from the interval $-1\leq z\leq 1$ to the entire plane. Explicitly, the few first harmonics for $z<1$ are 
\begin{subequations}
\bea
{\cal P}_0(q,\omega)&=&-1+\frac 12 z\left( \ln\frac{1+z}{1-z}-i\pi\right),\label{P03}\\
{\cal P}_1(q,\omega)&=&z\left[-1+\frac 12 z\left(\ln\frac{1+z}{1-z}-{i\pi}\right)\right],\label{P13}\\
{\cal P}_2(q,\omega)&=&\frac z2\left[-3z+\frac{3z^2-1}{2}\left(\ln\frac{1+z}{1-z}-i\pi\right)\right].\nn\\\label{P23}
\eea
\end{subequations}
Of course, ${\cal P}_0(q,\omega)$  coincides (up to a factor of the density of states) with the usual polarization bubble in the semiclassical limit of $q\ll\kf$ both in 2D and 3D.

Similarly, the Landau interaction function is expanded as
\bea
F^a(\hat\bk\cdot\hat\bk')=\left\{\begin{array}{ccc}\sum^\infty_{\ell=-\infty} e^{i\ell(\theta_\bk-\theta_{\bk'})}F_\ell^a,
\\
\sum_{\ell=0}^\infty (2\ell+1) P_\ell(\hat\bk\cdot\hat\bk')F_\ell^a.
\end{array}
\right.
\label{Fexp}
\eea

An expansion of ${\bf u}(\hat\bk;\bq,\omega)$ requires an additional consideration. In general, there are three independent vectors that $\bu$ may depend on: $\hat \bk$, $\bq$, and $\bf{B}$. However, since the magnetic field is transverse, i.e., $\bq\cdot\bf{B}=0$, there are in fact only two independent vectors. We can always choose $\bf{B}$ as the $z$-axis and t$\bq$ as the $x$-axis.  Then
the vector function $\bu$ depends only on the angle $\theta_\bk$ between $\hat\bk$ and $\bq$, while the direction of $\bf{B}$ defines  the direction of $\bu$. 
Therefore, ${\bf u}(\hat\bk;\bq,\omega)$
can be expanded over angular harmonics in the same way as ${\cal P}(\hat k;\bq,\omega)$ 
\bea
\bu(\hat k;\bq,\omega)=\left\{\begin{array}{ccc}\sum^\infty_{\ell=-\infty} e^{i\ell\theta_\bk}\bu_\ell(q,\omega),\\
\\
\sum_{\ell=0}^\infty (2\ell+1) P_\ell(\cos\theta_\bk)\bu_\ell(q,\omega).
\end{array}
\right.
\label{uexp}
\eea

Substituting Eqs.~(\ref{Pexp}), (\ref{Fexp}), and (\ref{uexp}) into Eq.~(\ref{ueq}), 
we obtain an infinite system 
of equations for  ${\bf u}_{\ell}(q,\omega)$. In 2D, this system reads
\bea
{\bf u}_{\ell}=-{\cal P}_{\ell} {\bf B}+\sum_{\ell'} {\cal P}_{\ell-\ell'}^{\phantom{a}}F^a_{\ell'}{\bf u}_{\ell'},
\label{system2D}
\eea
where we suppressed the argument $(q,\omega)$ for brevity.
Noting that $F^a_{-\ell}=F^a_\ell$ and ${\cal P}_{-\ell}={\cal P}_{\ell}$, we deduce that ${\bf u}_{-\ell}={\bf u}_{\ell}$. 

In 3D, the corresponding system of equations reads
\bea
\label{system3D}
\bu_\ell=-{\cal P}_\ell{\bf B}+\sum^\infty_{\ell',\ell''=0} (2\ell'+1)\begin{pmatrix}
\ell&\ell''&\ell'\\
0&0&0\end{pmatrix}^2{\cal P}_{\ell'}F^a_{\ell''} \bu_{\ell''},\nn\\
\eea
where $\begin{pmatrix}
j_1&j_2&j_3\\
m_1&m_2&m_3\end{pmatrix}$ is a $3j$-symbol. In deriving Eq.~(\ref{system3D}), we used the identity \cite{Landau:QM}
\bea
\int^{1}_{-1} \frac{dx}{2} P_{\ell_1}(x) P_{\ell_2}(x) P_{\ell_3}(x)=\begin{pmatrix}
\ell_1&\ell_2&\ell_3\\
0&0&0\end{pmatrix}^2\eea
and permutation symmetry of the last result.

The induced magnetization is related to the zeroth harmonic of $\bu$ via\beq
{\bf M}(q,\omega)=\text {Tr}\int\frac{d^Dk}{(2\pi)^D}\bs\delta\hat n_{\bk}(q,\omega)=N_F^*{\bf u}_0(q,\omega).
\label{M}
\eeq
Once Eqs.~(\ref{system2D}) and (\ref{system3D}) are solved, Eq.~(\ref{M}) allows one to read
off the expression for the spin susceptibility.

\subsubsection{Limiting cases}
\label{sec:limits}
\paragraph{Quasistatic limit.}
Equations (\ref{system2D}) and (\ref{system3D}) can be solved analytically for an arbitrary Landau function in the quasistatic regime, i.e., for $\omega\ll \vf^*q$ or, equivalently, for $z\ll 1$.
In 2D, ${\cal P}_\ell$ in 
Eq.~(\ref{Pl2Dexp}) is reduced in this limit to 
\beq
 {\cal P}_{\ell}=-\delta_{\ell,0}+(-i)^{\ell+1} z. \label{even}
\eeq
This asymptotic form can be readily reproduced by noting that for $z\ll 1$ the integral in Eq.~(\ref{Pl2D}) is controlled by the region where $\bk$ is almost perpendicular to $\bq$, i.e., where $\theta_\bk\approx \pm \pi/2$. If $\ell$ is odd, one can  safely set $\omega$ to zero in the denominator
of the integrand in Eq.~(\ref{Pl2D}) because the zeroes of $\cos(\ell\theta_\bk)$ and $\cos\theta_\bk$ at $\theta_\bk=\pm\pi/2$ cancel each other. Then ${\cal P}_{2n+1}$ is real and proportional to $z$ while
its 
imaginary part
occurs only to order $z^3$.
 (The smallness of $\im{\cal P}_{2n+1}$  implies the weakness of Landau damping in odd angular momentum channels, which is an important feature of a nematic FL. \cite{oganesyan:2001}) If $\ell$ is even, $\cos(\ell\theta_\bk)$ is finite at $\theta_\bk=\pm\pi/2$ but $\cos\theta_\bk$ vanishes, so the pole in the integrand needs to be circumvented, which gives a factor of $i\pi$. As a result,  the integral in Eq.~(\ref{Pl2D}) is purely imaginary and still proportional to $z$.  
Combining the  even and odd cases  together, we arrive at Eq.~(\ref{even}). 

Substituting Eq.~(\ref{even}) into Eq.~(\ref{system2D}) and solving the resulting system iteratively to first order in $z$, we find 
\bea
{\bf u}_{\ell}&=&{\bf B}(\bq,\omega)\left[\frac{\delta_{\ell0}}{1+F^a_0}
-(-i)^{\ell+1} z \frac{1}{(1+F_\ell^a)(1+F_0^a)}\right.\nn\\
&&\left.+{\cal O}(z^2)\right]
\eea
Substituting the last result with $\ell=0$ into Eq.~(\ref{M}), we obtain an asymptotic form of the spin susceptibility for $\omega\ll \vf^*q$\cite{nozieres,wu:2018}
\bea
\chi_s(\bq,\omega)=\frac{N_F^*}{1+F_0^a}\left[1+\frac{i\omega}{ v^*_Fq(1+F_0^a)}\right].\label{static}
\eea

In 3D, the asymptotic forms of ${\cal P}_\ell$ in Eq.~(\ref{Pl3Dexp}) for $\ell=2m$ and $\ell=2m+1$ read
\bea
{\cal P}_{2m}(q,\omega)&=&-\delta_{m,0}-\frac{i\pi}{2}\frac{(-)^m(2m)!}{(m!)^2}z,\nn\\
{\cal P}_{2m+1}(q,\omega)&=&(-)^{m+1}\frac{2^{m-1}}{m+\frac 12}\frac{m!}{(2m-1)!!}z.
\eea
The final result for the spin susceptibility in the 3D case differs from that in Eq.~(\ref{static}) only by a coefficient of $\pi/2$ in the imaginary part.

\paragraph{The region near the continuum boundary.}
Another region which can be analyzed for an arbitrary Landau function is just below the continuum threshold, defined by the condition $0<(\vf^*q-\omega)/\vf^*q=1-z\ll 1$.
We discuss the 2D case first. For $z\approx 1$, the integral for ${\cal P}_\ell(q,\omega)$ in Eq.~(\ref{Pl2D}) is controlled by a region of small  $\theta_\bk$. This means that most of the spectral weight comes from particle-hole pairs that are moving along $\bk$. Replacing $\cos\theta_\bk$ in the denominator of  Eq.~(\ref{Pl2D}) by $1-\theta^2_\bk/2$ and $\cos(\ell \theta_\bk)$  in the numerator by unity, extending the region of integration from $(0,\pi)$ to $(0,\infty)$, and solving the resultant integral, we find that ${\cal P}_\ell(q,\omega)$ in this approximation does not depend on $\ell$ and coincides with ${\cal P}_0(q,\omega)$:
\bea
{\cal P_\ell}(q,\omega)={\cal P}_0(q,\omega)
= -\frac{i}{\sqrt{2(1-z)}}.\label{Pt}
\eea
Taking ${\cal P_\ell}(q,\omega)$ out of the sum in Eq.~(\ref{system2D}), we solve the system by multiplying its both sides by $F^a_\ell$ and summing over $\ell$.
This yields the following limiting form of $\chi_s(q,\omega)$ near the threshold:
\bea
\im \chi_s(\bq,\omega)=N_F^*C \sqrt{\frac{2\left(\vf^*q-\omega\right)}{\vf^*q}}\theta(\vf^*q-\omega),\label{thr2D}
\eea
where 
$C=\frac{1}{\left(\sum_\ell F^a_\ell\right)^2}$. Note that  $\sum_\ell F^a_\ell$ in 2D is equal to the Landau function in the forward-scattering limit, i.e., 
at $\theta=0$. 

In 3D, the leading singularity in ${\cal P}_\ell$  can be obtained by integrating Eq.~(\ref{Pl3D}) by parts. This yields
\bea
{\cal P_\ell}(q,\omega)={\cal P}_0(q,\omega)=-\frac 12 \ln(1-z).
\eea
As in the 2D case, we can take ${\cal P}_\ell$ outside the sum in Eq.~(\ref{system3D}). The $3j$-symbol is eliminated with the help of the normalization condition\cite{balcar}
 $\sum_{\ell'}(2\ell'+1)\begin{pmatrix}
\ell&\ell''&\ell'\\
0&0&0\end{pmatrix}^2=1$, and the resultant system is solved in the same way as in 2D.  The threshold behavior of $\im\chi_s(q,\omega)$ in 3D is then  found to be
\bea
\im\chi_s(q,\omega)=N_F^*C\frac{2\pi}{\ln^2\frac{2\vf^*q}{\vf^*q-\omega}},\label{thr3D}
\eea
 where $C$ is given right after Eq~(\ref{thr2D}).

However,  a more detailed analysis shows that while $\im \chi_s(\bq,\omega)$ indeed vanishes near the threshold as $\sqrt{\vf^*q-\omega}$ and $1/\ln^2(\vf^*q-\omega)$ in 2D and 3D, respectively,
the overall prefactor contains a more complicated combination of $F_\ell^a$ compared to what is given after Eq.~(\ref{thr2D}). The reason for this discrepancy is that the leading terms in ${\cal P}_\ell$ cancel each other and one needs to keep the subleading terms.  Finding a general form of $C$ in 
 Eqs.~(\ref{thr2D}) and (\ref{thr3D}) turns out to be a rather complicated problem which we are not going to address here. In the next section, we will derive explicit forms of $C$ for special models of the Landau function.
  
    \begin{figure}[tbh]
\includegraphics[width=1.0 \linewidth]
{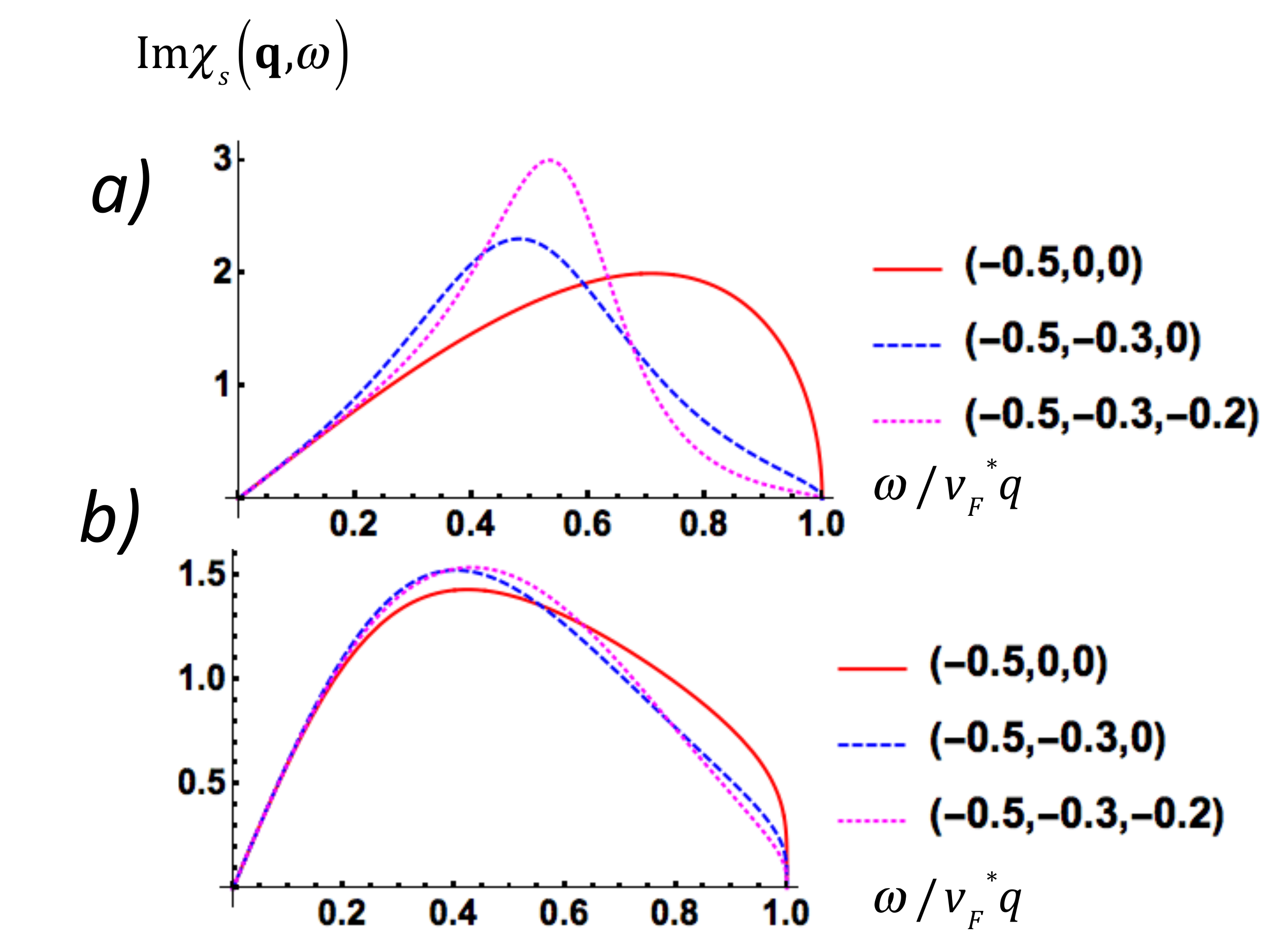}
\vspace{-0.2in}
\caption{Imaginary part of the spin susceptibility (normalized to its static value) for a 2D ({\em a}) and 3D ({\em b}) Fermi liquid, obtained by solving the kinetic equation for a model form of the Landau function, which contains up to three first harmonics in the spin channel. The legends specify the harmonics of the Landau function following the $(F_0^a,F_1^a,F^a_2)$ format. \label{fig:chif0f1}}
\end{figure}

\subsubsection{Special cases of the Landau function}
\label{sec:special}
Aside from the two limits considered above, the dynamical spin susceptibility can be found in a closed form only if  the Landau function contains a finite number of harmonics.
 
The simplest model is the $s$-wave approximation: $F^a(\theta)=F^a_0$. In this case, Eqs.~(\ref{system2D}) and (\ref{system2D})  are reduced 
to the same form 
\bea
\bu_l=-{\cal P}_\ell{\bf B}+F_0^a\bu_0{\cal P}_{\ell}\label{system0}
\eea
To arrive at this result in 3D one needs to recall that 
\bea
\begin{pmatrix}
\ell&0&\ell'\\
0&0&0\end{pmatrix}=(-)^{\ell'}\delta_{\ell',\ell'}\frac{1}{\sqrt{2l+1}}.\label{3j_1}
\eea
Setting $\ell=0$ in Eq.~(\ref{system0}) and using Eq.~(\ref{M}), we immediately obtain the familiar RPA result
\bea
\chi_s(\bq,\omega)=-N_F^*\frac{{\cal P}_0}{1-{\cal P}_0F^a_0},
\label{swave}
\eea
where ${\cal P}_0$ is given by Eqs.~(\ref{P02}) and (\ref{P03}) in 2D and 3D, respectively.
Equation (\ref{swave}) is equivalent to Eq.~(\ref{hubbard}) upon identifying $F_0^a$ with $-N_FU/2$.
The imaginary parts of $\chi_s(\bq,\omega)$ for $F_0^a=-0.5$ in 2D and 3D are shown by solid lines in panels {\em a} and {\em b} of Fig.~\ref{fig:chif0f1}, respectively.

Next, we assume that the Landau function contains two harmonics: $F_0^a$ and $F_1^a$.  
In this case, 
an infinite system for the harmonics of the distribution function in 2D [Eq.~(\ref{system2D})] is reduced to a $2\times 2$ system for ${\bf u}_0$ and ${\bf u}_{1}={\bf u}_{-1}$:
\bea
\bu_0&=&-{\cal P}_0{\bf B}+{\cal P}_0F^a_0\bu_0+2{\cal P}_1F^a_1\bu_1\nn\\
\bu_1&=&-{\cal P}_1{\bf B}+{\cal P}_1F^a_0\bu_0+({\cal P}_0+{\cal P}_2)F^a_1\bu_1.
\eea
Solving this system and substituting ${\bf u}_0$ into  Eq.~(\ref{M}), we obtain for the spin susceptibility
\begin{widetext}
\bea
\chi_s(\bq,\omega)=-N_F^*\frac{{\cal P}_0\left[1-F_1^a\left({\cal P}_0+{\cal P}_2\right)\right]+2{\cal P}_1^2
F_1^a}
{\left(1-{\cal P}_0F_0^a\right)\left[1-F_1^a\left({\cal P}_0+{\cal P}_2\right)\right]
-2F^a_1F_0^a{\cal P}_1^2},
\label{f0f12D}
\eea
\end{widetext}
where ${\cal P}_{0\dots 2}$ are given by Eqs.~(\ref{P02}-\ref{P22}). [The same result has recently been derived in Ref.~\onlinecite{wu:2018} by resumming the ladder diagrams.] The imaginary part of Eq.~(\ref{f0f12D}) for $F_0^a=-0.5$ and $F_1^a=-0.3$ is shown by the dashed line in Fig.~\ref{fig:chif0f1}{\em a}. 
For $\omega\ll \vf^*q$ Eq.~(\ref{f0f12D}), reproduces the quasistatic limit, Eq.~(\ref{static}), as it should. 
Expanding  Eq.~(\ref{f0f12D}) near the continuum boundary ($\omega\approx \vf^*q$), we reproduce Eq.~(\ref{thr2D})
with 
\bea
C=\left(\frac{1+F_1^a}{F^a_0+2F_1^a+F_0^aF_1^a}\right)^2.\label{C2}
\eea
For this model of the Landau function,  $\sum_\ell F^a_\ell=F_0^a+2F_1^a$. Therefore  the actual result for $C$ in Eq.~(\ref{C2}) differs from the formula right after Eq.~(\ref{thr2D}), which was obtained by neglecting higher-order terms in ${\cal P}_\ell$.  

Finally, we consider the case of the Landau function with the  first three harmonics: $F^a_0,F_1^a$, and $F_2^a$. The analytic form of $\chi_s(\bq,\omega)$ for this case is too long to be displayed here; its imaginary part is shown by the dotted line in Fig.~\ref{fig:chif0f1}{\em a} for $F_0^a=-0.5$, $F_1^a=-0.3$, and  $F_2^a=-0.2$. Near the threshold, $\im\chi_s(\bq,\omega)$ is again reduced to Eq.~(\ref{thr2D}) with 
\bea
C=\frac{\left(1+F_1^a\right)^2\left(1+\left[F_2^a\right]^2\right)+2F_2\left(1+2\left[F_1^a\right]^2\right)}{F_0^a+2F_1^a+2F_2^a+F_0^a\left(F_1^a+F_2^a+F_1^aF_2^a\right)}.\nn\\
\eea
As in the previous case, $C$ is also not expressed entirely in terms of $\sum_\ell F^a_\ell$.

Comparing the three cases shown in Fig.~\ref{fig:chif0f1}{\em a}, we note that inclusion of higher harmonics has a rather strong effect on the overall shape of 
$\im\chi_s(\bq,\omega)$ in 2D. In particular, the square-root threshold singularity becomes less and less pronounced as the number of harmonics in the Landau function is increased.

Now we turn to the 3D case. For the Landau function containing the first two harmonics,  an infinite system of equations in Eq.~(\ref{system3D}) is reduced  to a $2\times 2$ one with the help of Eq.~(\ref{3j_1}) and another identity for the $3j$ symbols:\cite{balcar}
\bea
\begin{pmatrix}
1&1&\ell\\
0&0&0\end{pmatrix}&=\left\{\begin{array}{ccc}
\frac{\ell^2+\ell-2}{\sqrt{(2-\ell)!(\ell+3)!}},\;\text{for}\;0\leq\ell\leq 2\\
0;\text{otherwise}.
\end{array}
\right.
\eea
Using these identities, we obtain 
\bea
\bu_0&=&-{\cal P}_0{\bf B}+{\cal P}_0F^a_0\bu_0+{\cal P}_1F^a_1\bu_1\nn\\
\bu_1&=&-{\cal P}_1{\bf B}+{\cal P}_1F^a_0\bu_0+\frac 13\left({\cal P}_0+2{\cal P}_2\right)F^a_1\bu_1.\nn\\
\eea
Accordingly, for the spin susceptibility we find
\begin{widetext}
\bea
\chi_s(\bq,\omega)=-N_F^*\frac{{\cal P}_0\left[1-\frac 13 F_1^a\left({\cal P}_0+2{\cal P}_2\right)\right]+{\cal P}_1^2
F_1^a}
{\left(1-{\cal P}_0F_0^a\right)\left[1-\frac 13F_1^a\left({\cal P}_0+2{\cal P}_2\right)\right]
-F^a_1F_0^a{\cal P}_1^2},
\label{f0f13D}
\eea
\end{widetext}
where ${\cal P}_{0\dots 2}$ are given by Eqs.~(\ref{P03}-\ref{P23}).
Expanding Eq.~(\ref{f0f13D}) near the threshold, we reproduce Eq.~(\ref{thr3D}) with 
\bea
C=\frac{\left(1+\frac 13 F_1^a\right)^2}{\left(F_0^a+F_1^a+\frac 13 F_0^aF_1^a\right)^2}\label{C3D}
\eea
As in the 2D case, the actual form of $C$ differs from the formula obtained by neglecting higher-order terms in ${\cal P}_\ell$. 
The imaginary part of $\chi_s(\bq,\omega)$ in Eq.~(\ref{f0f13D}) is shown by the dashed line in Fig.~\ref{fig:chif0f1}{\em b} for $F_0^a=-0.5$, and $F_1^a=-0.3$. As in 2D, the analytic result for the Landau function with three harmonics is too long to be presented here; the corresponding imaginary part is shown by the dashed line in Fig.~\ref{fig:chif0f1}{\em b} for $F_0^a=-0.5$, $F_1^a=-0.3$, $F_1^a=-0.2$. We see that inclusion of higher harmonics in the 3D case has a less pronounced effect on the shape of $\chi_s(\bq,\omega)$ compared to the 2D one.
\subsubsection{Numerical solution of the kinetic equation}
\begin{figure}[tbh]
\includegraphics[width=1.0 \linewidth]{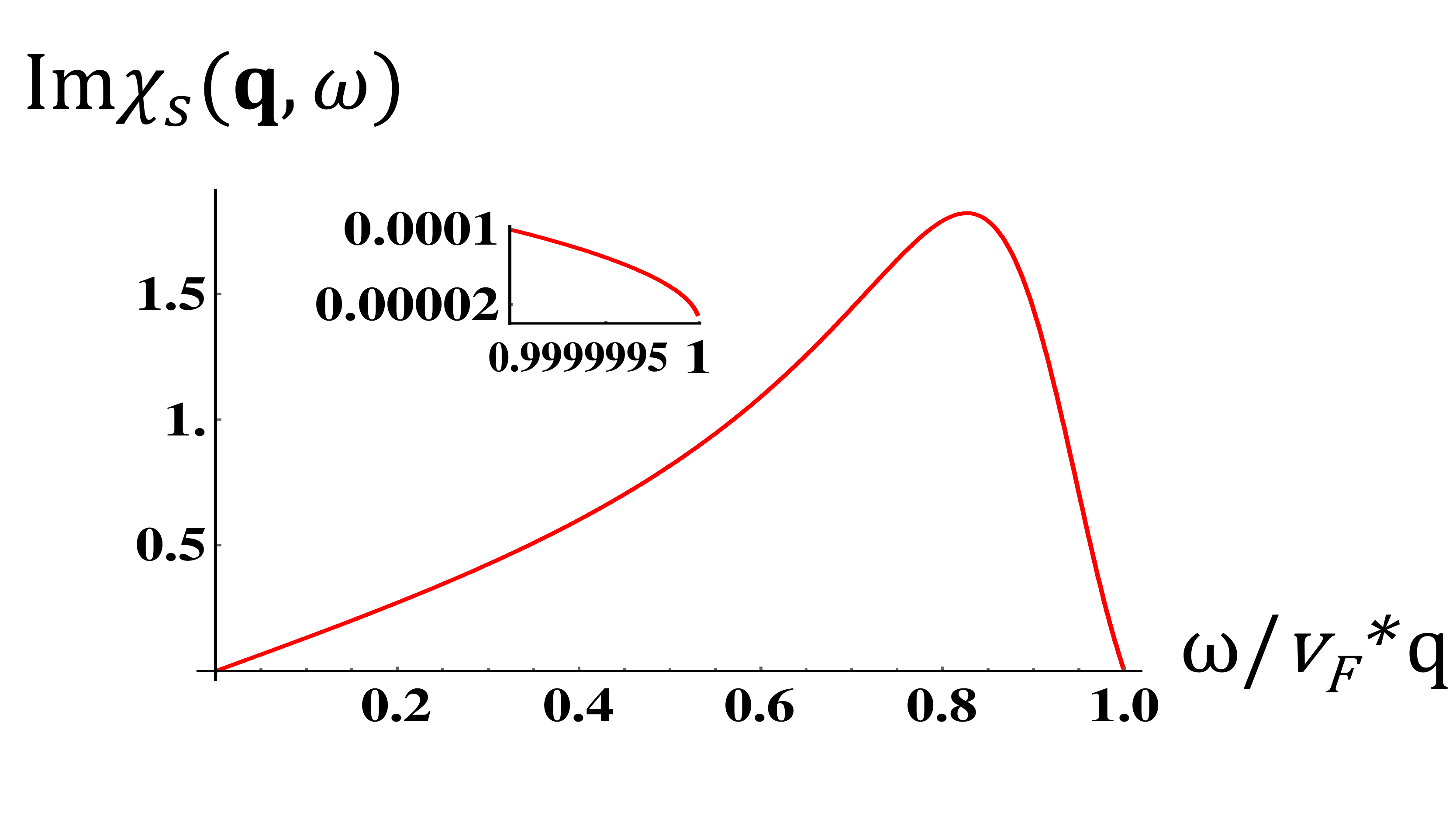}
\vspace{-0.2in}
\caption{Imaginary part of the spin susceptibility (normalized to its static value) for a model form of the Landau function corresponding to the statically screened Coulomb potential in 2D.  The system of equations (\ref{system2D}) was solved numerically using the first $101$ harmonics of the Landau function in Eq.~(\ref{LFC}) with $a=0.5$. \label{fig:KEC}}
\end{figure}

In this section, we present a numerical solution of Eq.~(\ref{system2D}) for a particular model of the Landau function corresponding to the statically screened Coulomb potential. To first order in such interaction and in 2D,\cite{chubukov:2010}
\bea
F^a(\theta)=-\frac 12 \frac{a}{\left\vert\sin\frac{\theta}{2}\right\vert+a},\label{LFC}
\eea
where $a=\kappa/2k_{\text{F}} $, $\kappa=2\pi e^2\nu_2$ is the inverse screening radius, and $0\leq \theta\leq 2\pi$.
The first $N$ harmonics of $F^a(\theta)$ are found numerically and then Eq.~(\ref{system2D}) is numerically diagonalized. The resultant
imaginary part of $\im\chi_s(\bq,\omega)$ is shown in Fig.~\ref{fig:KEC} for  $a=0.5$ and $N=101$. Note that the square-root threshold singularity of  $\im\chi_s(\bq,\omega)$ is not visible in the main panel of Fig.~\ref{fig:KEC}. This is the same behavior  as we have already seen for the Landau function with the first few non-zero harmonics [cf.  dashed and dotted lines in Eq.~\ref{fig:chif0f1}{\em a}]. However, the inset in Fig.~\ref{fig:KEC} shows that the  singularity is still present in a very narrow vicinity of the threshold.

\begin{figure}[t]
\includegraphics[width=1.0 \linewidth]
{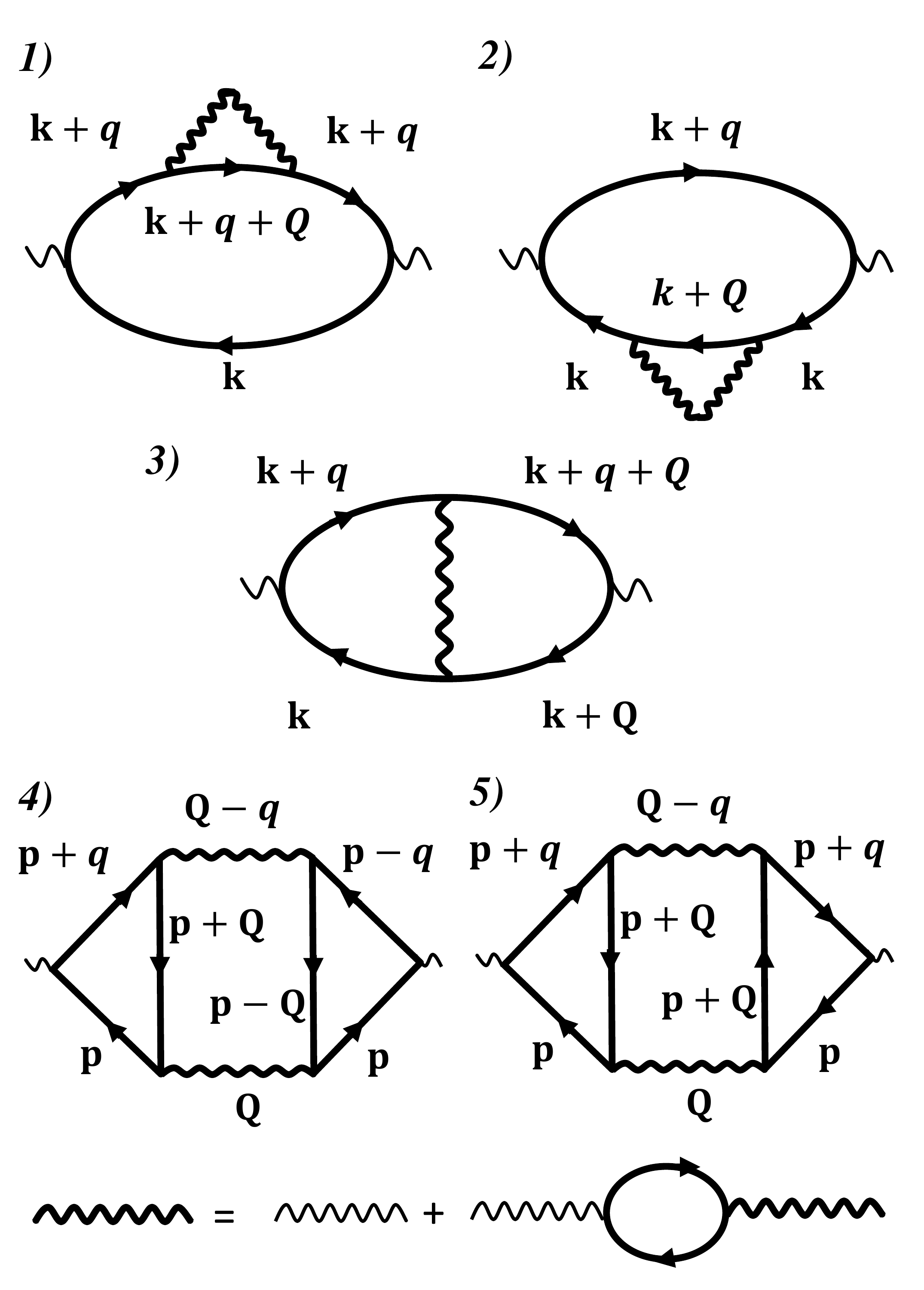}
\caption{Diagrams for a susceptibility to lowest order in the dynamically screened Coulomb interaction (wavy line). Diagrams 1-3 contribute give a correction to the spin susceptibility. 
 Diagrams 1-5 give a correction to the irreducible charge susceptibility. The full charge susceptibility is obtained by resumming  irreducible parts connected by a single interaction line, with the result given by Eq.~(\ref{charge}). \label{fig1}}
\end{figure}

\subsection{Charge susceptibility}
For a neutral FL, the expression for the compressibility  is derived along the same lines as in Sec.~\ref{sec:FLspin} except for an external perturbation in Eq.~(\ref{ke}) is replaced a classical force,
$-\boldsymbol{\nabla} U$, where $U$ is the potential energy, and the corresponding change in the occupation number is parameterized by a scalar rather than a vector function. Explicit results for compressibility is  obtained from those derived for the spin susceptibility in Secs.~\ref{sec:limits} and \ref{sec:special} simply by replacing the Landau parameters in the spin channel by those in the charge channel, i.e., $F_\ell^{a}\to F_{\ell}^{s}$, $\ell=0,1,2\dots$. 

For a charged FL, one needs to take into account the difference between the external and total electric fields acting on a given quasiparticle. The charge susceptibility can be related to the compressibility of a fictitious neutral FL with the same Landau function.\cite{nozieres} In the diagrammatic language, the compressibility of a neutral FL plays the role of an irreducible charge susceptibility, $\chi^{\text{irr}}_c(\bq,\omega)$, which contains all diagrams that cannot be separated into by cutting only one interaction line. The exact relation between the full and irreducible susceptibilities is \cite{nozieres}
\bea
\chi_c(\bq,\omega)=\frac{\chi^{\text{irr}}_{c}(\bq,\omega)}{1+U_0(\bq)\chi^{\text{irr}}_{c}(\bq,\omega)}.
\label{charge}
\eea

\section{Interacting quasiparticles}
\label{sec:any}
\subsection{Model}
\label{sec:form}

Interaction between quasiparticles can be taken into account by adding a collision integral to the right-hand side of the kinetic equation, Eq.~(\ref{ke}). Alternatively, one can take the same effect into account  by calculating diagrams for the susceptibilities {\em beyond} the RPA level for a particular model of {\it retarded}, i.e., dynamically screened interaction. In this paper, we adopt the diagrammatic method for the dynamically screened Coulomb interaction, 
\bea U({\bf q},\omega_{m}) = \left[U_0^{-1}({\bf q}) + \chi^{(0)}({\bf q},\omega_{m})\right]^{-1},\label{coulomb}\eea
where, as before, $\chi^{(0)}({\bf q},\omega_{m})$ is the free-electron susceptibility, $U_0(\bq)=2\pi e^2/q$ and $U_0(\bq)=4\pi e^2/q^2$ in 2D and 3D, correspondingly.

The one-loop diagrams for the polarization bubble are shown in Fig.~\ref{fig1}, where  the thick wavy lines denote the dynamically screened Coulomb potential, given by Eq.~(\ref{coulomb}). The vertices are given by unities in the charge channel and by the Pauli matrices in the spin one. The AL diagrams (diagrams 4 and 5) vanish identically in the spin channel due to spin traces, and the correction to the spin susceptibility is given by the sum of diagrams 1-3:
\bea
\delta\chi_s(\bq,\omega)=\sum_{\alpha=1\dots 3} \delta\chi^{(\alpha)}(\bq,\omega).\label{delta_chi_s}\eea
On the other hand, all diagrams contribute to the irreducible charge susceptibility \bea
\delta\chi^{\text{irr}}_{c}(\bq,\omega)=\sum_{\alpha=1\dots 5}\delta\chi^{(\alpha)}(\bq,\omega).
\eea
At the RPA level, $\delta\chi^{\text{irr}}_{c}(\bq,\omega)=0$ and $\chi_c(\bq,\omega)$ is reduced to Eq.~(\ref{chi_charge}).
To obtain a correction to the RPA result, we substitute $\chi^{\text{irr}}_c(q,\omega)=\chi^{(0)}(\bq,\omega)+\delta\chi^{\text{irr}}_{c}(\bq,\omega)$ into Eq.~(\ref{chi_charge}) and expand to 
lowest order in $\delta\chi^{\text{irr}}_{c}(\bq,\omega)$ to obtain
\bea
 \delta\chi_c(\bq,\omega)=\frac{\delta\chi^{\text{irr}}(\bq,\omega)}{\left[1+U_0(\bq)\chi^{(0)}(\bq,\omega)\right]^2}.\label{chi_c_RPA}`
\eea
\subsection{Spin susceptibility}
\label{sec:spin}


We note that only a dynamical interaction leads to damping of quasiparticles and thus can
give rise to a non-zero spectral weight of the susceptibility above the continuum boundary. Therefore, it is convenient to subtract off the static part of the interaction 
\bea
U(\bQ,\Omega_l)&=&U(\bQ,\Omega_l)-U(\bQ,0)+U(\bQ,0)\nn\\
&&\equiv U^{\text{dyn}}(\bQ,\Omega_l)+U(\bQ,0).
\eea
The contribution from the static part has been effectively accounted for in Sec.~\ref{sec:rpa} by solving
the FL kinetic equation without a collision integral. All one needs to do is to calculate the Landau function to the lowest order in  $U(\bQ,0)$. The result will be some insignificant modification
of the spectral weight below the continuum boundary. In what follows, we neglect this contribution and focus on the one from the dynamic part of the interaction, $U^{\text{dyn}}(\bQ,\Omega_l)$.

After some manipulations with the Green functions, the sum of diagrams 1-3 in Fig.~\ref{fig1}, which give a correction to the spin susceptibility [Eq.~(\ref{delta_chi_s})], can be written compactly as (see Appendix  \ref{sec:appb} for details) \bwt
\bea
\label{chis}
\delta\chi_s(\bq,\omega_m)
=-\int \int \int \int &&\frac{d^DQ d^Dkd\Omega_ld\varepsilon_n}{(2\pi)^{2(D+1)}}
U^{\text{dyn}}(\bQ,\Omega_l)\frac{
\left(\epsilon_{\bk+\bq}-\epsilon_\bk-\epsilon_{\bk+\bQ+\bq}+\epsilon_{\bk+\bQ}\right
)^2}{(i \omega_m - \epsilon_{\bk+\bQ+\bq} + \epsilon_{\bk+\bQ})^2 (i \omega_m - \epsilon_{\bk+\bq}+ \epsilon_\bk)^2}\nn\\
&&\times \left[G(\bk,\e_n)- G(\bk+\bq,\e_n+\omega_m)\right] \left[G(\bk+\bQ,\e_n+\Omega_l)- G(\bk+\bQ+\bq,\e_n+\Omega_l+\omega_m)\right]\nn\\
\eea
\ewt
Equation (\ref{chis}) is valid for $q\ll k_{\text{F}} $ and arbitrary $\omega$. For a parabolic spectrum ($\epsilon_\bk=k^2/2m-k_{\text{F}} ^2/2m$), the combination $(\epsilon_{\bk+\bq}-\epsilon_\bk-\epsilon_{\bk+\bQ+\bq}+\epsilon_{\bk+\bQ})$ is reduced to $(\bq\cdot\bQ)^2$ for arbitrary $q$ and $Q$. For an arbitrary spectrum, this combination
is simplified to $\left[({\bv}_{\bk+\bQ}-{\bf v}_{\bk})\cdot\bq\right]^2$ for small $q$.
\subsubsection{Low frequencies: $\omega\ll \vf\kappa$}
In this regime, typical momentum transfers $Q$ are either logarithmically larger (in 2D) or on the order (in 3D) of the inverse screening radius $\kappa$, which needs to be chosen much smaller than $k_{\text{F}} $ to keep the perturbation theory under control. On the other hand, the internal bosonic frequencies are on the order of the external one: $\Omega\sim \omega$. Therefore, for external frequencies in the range $\omega\ll \vf\kappa\ll E_{\text{F}}$, the dynamical polarization bubble in the screened Coulomb potential can be expanded to leading order in $\Omega/\vf Q$: $\Pi^{(0)}(\bQ,\Omega)\approx -N_F(1+iC_D\Omega/\vf Q)$, where $C_2=1$ and $C_3=\pi/2$.  Note that $\vf\kappa$ is on order of the plasma frequency in 3D and of the plasmon dispersion evaluated at $q\sim\kappa$ in 2D.
The range of $\omega$ specified above corresponds to the FL regime, in which the imaginary part of the self-energy scales as $\max\{\omega^2,T^2\}$.\cite{agd:1963} 

The rest of the calculations is fairly straightforward (see Appendix \ref{sec:se+l_2D} for details). The final result valid for an arbitrary ratio $\omega/\vf q$ and for parabolic single-particle dispersion is given by
\bea
\delta\chi_s(\bq,\omega)
=
 \lambda_{D}
\times
\left\{ 
\begin{array}{ccc}
\frac{q^2 \omega^4}{\left(\vf^2q^2-\omega^2-i \delta\text{sgn}\omega \right)^{5/2}}\ln\frac{E_{\text{F}} }{
\vf\kappa,
}\\
\\
i\frac{k_{\text{F}} q^2 \omega^3}{\left[  v^2_Fq^2 -\omega^2\right]^2},
\end{array}
\right.
\label{chi_s_main}
\end{eqnarray}
where the dimensionless coupling constants are given by
\begin{subequations}
\bea
\lambda_{2} &=& \frac{e^4}{6\pi^2 \vf^2}
= \frac{r_s^2}{
12
\pi^2}
\approx 
8.0
\times 10^{-3} r_s^2,
\label{lambda2}\\
\lambda_{3} &=& \frac{e^4 }{36\pi^2 v^2_F}\frac{k_{\text{F}} }{\kappa}= \frac{r_s^{3/2}}{108\pi^2}
\approx 9.4\times 10^{-4} r_s^{3/2},\label{lambda3}
\eea
\end{subequations}
where $r_s$ is the standard dimensionless coupling constant for the Coulomb interaction, equal to the average distance between electrons measured in units of the Bohr radius.
The first (second) lines in Eqs.~(\ref{chi_s_main}) and (\ref{lambda2},\ref{lambda3}) refer to the 2D (3D) case.  Note that the numerical prefactors  in both cases are quite  small.

We emphasize that although the results in Eq.~(\ref{chi_s_main}) were derived in a particular model of a dynamically screened Coulomb interaction, they are expected to apply to any generic FL. The only change will be in the particular form of the prefactor $\lambda_D$. Indeed,
Eq.~(\ref{chi_s_main}) resulted from the Landau-damped form of the dynamic interaction, which is expected to be obeyed in any FL.


We see that $\im\delta \chi_s(\bq,\omega)$ is non-zero outside the continuum ($|\omega|>\vf q$): this is the main difference compared to the result obtained for noninteracting quasiparticles. Explicitly,
\bea
&&\im\delta\chi_s(\bq,\omega)
= \lambda_{D}\nn\\
&&\times
\left\{ 
\begin{array}{ccc}
\frac{q^2 \omega^4\text{sgn}\omega}{\left(\omega^2-\vf^2q^2\right)^{5/2}}\ln\frac{E_{\text{F}} }{
\vf\kappa
}\theta(|\omega|-\vf q),\\
\frac{k_{\text{F}} q^2 \omega^3}{\left(\omega^2- v^2_Fq^2 \right)^2}.
\end{array}
\right.
\label{im_chi_s_main}
\end{eqnarray}
In the 2D case,  the residual interaction between quasiparticles does not affect the spectral weight below the continuum boundary, because $\delta\chi_s^{\text{dyn}}(\bq,\omega)$ is purely real for $|\omega|<\vf q$. On the other hand, the corresponding correction in 3D is purely imaginary, which means that the real part of susceptibility is not affected by the residual interaction between quasiparticles.

Far away from the continuum boundary ($\vf q\ll |\omega|\ll \vf\kappa$), $\im\delta\chi_s$ assumes a universal form
\bea
\im\delta \chi_s(\bq,\omega)=\lambda_Dk_{\text{F}} ^{D-2}\frac{q^2}{\omega},\label{tail}
\eea 
with an extra factor of $\ln (k_{\text{F}} / \kappa)\sim \ln r^{-1}_s$ in  
2D. 

On approaching the continuum boundary ($|\omega|\to \vf q$),  $\im\delta \chi_s(\bq,\omega)$ diverges as 
\begin{eqnarray}\label{result_any_b}
\im\delta\chi_s
({\bf q}, \omega) 
&=& \lambda_{D}\times
\left\{ 
\begin{array}{ccc}
&&\frac{\sqrt{2}}{8}\frac{q^2 (\vf q)^{3/2}}{\left(\omega-\vf q\right)^{5/2}}\theta(|\omega|-\vf q),\\
\\
&&\frac{1}{4}\frac{k_{\text{F}} \vf q^3}{\left(\omega -\vf q\right)^2}.
\end{array}
\right.
\end{eqnarray}

To eliminate the threshold singularities, one needs to resum the series for the spin susceptibility. The lowest-order term is the irreducible susceptibility found in this section.
The next-order term contains two irreducible susceptibilities, {\em not} yet integrated over the angle between $\bk$ and $\bq$, which are separated by an irreducible static vertex proportional to the Landau function in the spin channel. The second-order term contains three irreducible susceptibilities and  two irreducible vertices, etc. The series can be cast into the form of an integral equation which cannot be solved analytically for a general form of the Landau function. Approximating the Landau function by the $\ell=0$ harmonic and expanding to the lowest order in the irreducible susceptibility, we obtain a familiar RPA result\cite{ma:1968}
\bea
\label{res_rpa}
\im\delta\tilde\chi_s(\bq,\omega)=\frac{\delta\im\chi_s(\bq,\omega)}{\left[1-U\chi^0_s(\bq,\omega)/2\right]^2},
\eea
where $\im\delta\chi_s(\bq,\omega)$ is the imaginary part of the irreducible susceptibility given by Eq.~(\ref{im_chi_s_main}), $\chi^0_s(\bq,\omega)$ is the free-electron spin susceptibility, and $U=-2F_0^a/N_F$. Of course, such an approach is not rigorous, because $\im\delta\chi_s(\bq,\omega)$ was calculated for a long-range Coulomb potential while we approximated the interaction by a delta-function, when resumming the RPA series. Nevertheless, it does give an idea of how the threshold singularities in $\im\delta\chi_s(\bq,\omega)$ are weakened due to concomitant divergences in $\chi^0_s(\bq,\omega)$.  For example, the $(\omega-\vf q)^{-5/2}$ singularity in Eq.~(\ref{result_any_b}) is reduced to the $(\omega-\vf q)^{-3/2}$ singularity of the resummed susceptibility in Eq.~(\ref{res_rpa}). Complete elimination of the threshold singularities requires additional resummation of the series, which we will not attempt here.

The results presented above can be readily generalized for an isotropic but otherwise arbitrary single-particle dispersion, $\epsilon_\bk=\epsilon_k$. The only changes will be in the values of the dimensionless coupling constants in Eq.~(\ref{chi_s_main}). For example, it can be readily shown that the coupling constant in 2D [$\lambda_2$ in Eq.~(\ref{lambda2})] needs to be replaced by
\bea
\bar\lambda_2=\frac{e^4}{6\vf^2} \left(\frac{N_F}{m_{\text{eff}}}\right)^2,
\eea
where $\vf=d\epsilon_k/dk\vert_{k=k_F}$ is the group velocity, $N_F=k dk/\pi d\epsilon_k\vert_{k=k_F}$ is the density of states at the Fermi level, and the effective mass $\bar m$ is defined as
\bea
\frac{1}{m^2_{\text{eff}}}=\frac{1}{m^{*2}}+\frac{1}{2\tilde m^{2}}+\frac{1}{m^*\tilde m}
\eea
with
\bea
\frac{1}{m^*}=\frac{1}{k}\frac{d\epsilon_k}{dk}\Big\vert_{\kf}\;
\text{and}\;\frac{1}{\tilde m}= k\frac{d}{dk}\left(\frac{1}{k}\frac{d\epsilon_k}{dk}\right)\Big\vert_{\kf}.
\eea
For parabolic dispersion, $1/\tilde m=0$ while $m_{\text{eff}}=m^*=m$, and we recover Eq.~(\ref{lambda2}). 
\subsubsection{Higher frequencies and the f-sum rule}
\label{sec:fsum}
At the level of non-interacting quasiparticles, the only region that contributes to 
the spin $f$-sum rule 
\bea
\int^\infty_0 \frac{d\omega}{\pi}\omega\im\chi_s(\bq,\omega)=\frac{n_0q^2}{2m}\label{fsum}
\eea 
is the particle-hole continuum, $0\leq \omega\leq \vf q$ (here, $n_0$ is the number density). Due to residual interaction between quasiparticles, the spectral weight ``leaks out'' from the continuum, and now Eq.~(\ref{fsum}) needs to be satisfied over the whole range of frequencies. 

In the previous section, we found that $\im\delta\chi_s(\bq,\omega)\propto q^2/\omega$ for 
$\vf q\ll \omega\ll \vf\kappa$. A slow, $1/\omega$ decrease of $\im\delta\chi_s(\bq,\omega)$ in this range  of frequencies is insufficient to guarantee that the spin $f$-sum rule 
is satisfied. Therefore, we need to consider higher frequencies, $\omega\gg \vf\kappa$. In its turn, this interval can be separated into two:
$\vf\kappa\ll\omega\ll\Ef$ (intermediate frequencies) and $\omega\gg\Ef$ (high frequencies).
The behavior in these intervals differ substantially between 2D and 3D, and we discuss these two cases separately.

\paragraph{2D.} For $\vf\kappa\ll\omega\ll\Ef$, the only change compared to the case of $\omega\ll\vf\kappa$  is that the logarithmic integral over the momentum transfer $Q$ is now to be cut at $Q\sim|\omega_m|/\vf$ rather than at $Q\sim \kappa$. As a result, the factor of $\ln (\kf/\kappa)$ in the top line of Eq.~(\ref{chi_s_main}) is replaced by
 $\ln (\Ef/|\omega_m|)$. After analytic continuation, we find 
  \bea
 \im\delta\chi_s(\bq,\omega)=\lambda_2\frac{q^2}{\omega}\ln\frac{\Ef}{|\omega|}.\label{2di}
 \eea
We see that the decay of  $ \im\chi_s(\bq,\omega)$ 
 is even slower than at lower frequencies and, therefore, we need to consider  the range of $\omega\gg \Ef$ in order to make sure that the $f$-sum rule is satisfied. Details of the calculation are presented in Appendix {\ref{app:s_interm}}; the final result for this range reads
\bea
\im\delta\chi_s(\bq,\omega)=3\pi^2\lambda_2 \frac{q^2\Ef^3}{\omega^4}\text{sgn}\omega.
\label{2dh}
\eea
A fast, $1/\omega^4$ decay guarantees the convergence of the integral in Eq.~(\ref{fsum}).
We conclude that in 2D
the $f$-sum rule for the spectral weight above the continuum is satisfied at $\omega\sim \Ef$.
\paragraph{3D.}
In 3D, the integral over the momentum transfers is not logarithmic [an indication of which is the lack of the $\ln r_s$ factor in the corresponding result for $\omega\ll \vf\kappa$, Eq.~(\ref{chi_s_main})]. Therefore, in contrast to the 2D case, the behavior of $\im\delta\chi_s(\bq,\omega)$ changes dramatically already in the intermediate frequency range ($\vf\kappa\ll\omega\ll\Ef$). As shown in Appendix \ref{app:spin_high}, in this range we have
\bea
\im\delta\chi_s(\bq,\omega)=\frac{8\ln 2}{3\pi^2} \frac{e^4}{\vf^2} k_F\frac{q^2\Ef}{\omega^2}\text{sgn}\omega.\label{3di}
\eea
The integral in Eq.~(\ref{fsum}) still diverges but now only logarithmically. This means that the spectral weight above the continuum is distributed over a (formally) broad interval between $\vf\kappa$ and $\Ef$. One should expect even faster decay at $\omega\ll\Ef$; indeed, we estimate that $\im\delta\chi_s(\bq,\omega)\propto \text{sgn}\omega/|\omega|^{5/2}$ in this range.

Note that the condition $\vf\kappa\ll \Ef$ can be satisfied only at weak coupling; in most of real systems, $\vf\kappa\gtrsim\Ef$. Therefore, it would be correct to say that the spectral weight above the continuum comes from the region $\omega\sim \Ef$ both in 2D and 3D. Also note that, unlike the low-frequency form [Eq.~(\ref{chi_s_main})], the asymptotic forms at higher frequencies [Eqs.~(\ref{2di}), (\ref{2dh}), and (\ref{3di})] are not supposed to be universal but rather specific for a given model of the interaction.
\subsection{Charge susceptibility}
\label{sec:charge}
\subsubsection{Cancellation of diagrams for the irreducible part}
At the level of non-interacting quasiparticles,  the irreducible part of the charge susceptibility
 coincides with the spin susceptibility upon replacing the FL parameters in the spin sector  by those in the charge sector: $F_0^a\to F_0^s$, $F_1^a\to F_1^c\dots$.  
However, this one-to-one correspondence is lost once the interaction between quasiparticles is taken into account. Technically, the difference between the two channels is due to the AL diagrams (4 and 5 in Fig.~\ref{fig1}), which vanish in the spin channel due to tracing out the  Pauli matrices at the vertices to zero, but are non-zero in the charge channel. Note 
that the same AL  diagrams are responsible for the differences in the nonanalytic corrections in the charge and spin channels: such correction are absent in the former but present in the latter.\cite{belitz:1997}

The sum of the two AL diagrams can be written as
\bwt
\bea
\delta 
\chi_{\mathrm{AL}}(\bq,\omega_m)=
\delta\chi^{(4)}(\bq,\omega_m)+\delta\chi
^{(5)}(\bq,\omega_m)& =& 
4
\int \int \frac{d^DQd\Omega_l}{(2\pi)^{(D+1)}}
\left( {\cal T}^2 + \vert {\cal T}\vert^2 \right)U({\bf Q},\Omega)U({\bf Q} + {\bf q},\Omega_{l} + \omega_{m}),
\eea
where
\bea
{\cal T}\equiv
\int\int \frac{d^Dkd\e_n}{(2\pi)^{D+1}}G(\bp,\e_n)G(\bp+\bq,\e_n+\omega_m)G(\bp+{\bf Q},\e_n+\Omega_l).
\eea
\ewt
This expression 
is quite cumbersome for a generic ratio of $\omega$ to $\vf q$, so we restrict our analysis to the region of frequencies well above the continuum but below the energy scale set by the interaction, i.e., $\vf q\ll \omega\ll\vf\kappa$. 
After some algebra, the leading  term can be shown to be (see Appendix \ref{sec:AL} for details) 
\begin{align}
\label{tail}
\im\delta\chi_{\mathrm{AL}}
(\bq,\omega) = - 
\lambda_D\frac{q^2}{\omega}\times
\left\{
\begin{array}{ccc}
\ln(r_s^{-1}),\\
1,
\end{array}
\right.
\end{align}
where $\lambda_D$ are given by Eqs.~(\ref{lambda2}) and (\ref{lambda3}) in 2D and 3D, respectively. The AL is equal in magnitude and opposite in sign to
the contributions of the self-energy and ladder diagrams (diagrams 1-3 in Fig.~\ref{fig1}), which are the same  in the spin and charge channels.
Therefore the leading, $q^2/\omega$, term in the irreducible charge susceptibility is canceled out  between all the five diagrams
\bea\im\chi^{\text{irr}}_c(\bq,\omega)=\im\delta\chi_s(\bq,\omega)+\im\delta\chi_{\mathrm{AL}}=0\times {\cal O}(q^2/\omega)+\dots\nn\\
\eea

We stress that this cancellation is specific
feature of a Galilean-invariant system. Indeed, the self-energy and ladder diagrams are not crucially sensitive to particle-hole asymmetry, i.e., one still gets a non-zero result if the spectrum is linearized near the Fermi energy.  In contrast, the AL diagrams vanish if the spectrum is linearized, \cite{kamenev:1995} in which case the system becomes particle-hole symmetric. To get a non-zero result, one needs to break the particle-hole symmetry by retaining higher-order terms in the dispersion. This implies that the first three and the last two diagrams in Fig.~\ref{fig1} contain different parameters characterizing the single-particle dispersion, and, in general, cannot cancel each other. That such cancellation occurs in the Galilean-invariant case, i.e., for $k^2/2m$ dispersion, is not an accident but a consequence of a general relation between the charge susceptibility and longitudinal conductivity, discussed in the next section.

\subsubsection{Relation between the charge susceptibility and longitudinal conductivity}
Extracting the next after the $q^2$ term directly from diagrams 1-5 in Fig.~\ref{fig1} would be a difficult task. Fortunately, this problem can be circumvented by invoking a general relation between the charge susceptibility and the longitudinal conductivity, which is based on the Poisson equation
and Ohm's law.  In $D$-dimensions,
this relation reads
\bea
\chi_c(\bq,\omega)=\frac{iq^2\sigma_{||}(\bq,\omega)}{e^2\omega}\frac{1}{1+2\pi i A_D\sigma_{||}(\bq,\omega)/\omega},\label{fullcharge}
\eea
where $A_2=q$ and $A_3=2$. That the form of the equation above differs between the 2D and 3D cases is related to the fact that the units of the conductivity are different in different dimensions. However, Eqs.~(\ref{charge}) and (\ref{fullcharge}) show that the corresponding relation between the irreducible part of $\chi_c(\bq,\omega)$ and 
$\sigma_{||}(\bq,\omega)$ is independent of $D$:
\bea
\chi^{\text{irr}}_c(\bq,\omega)=\frac{iq^2}{e^2\omega}\sigma_{||}(\bq,\omega)
\eea
or
\bea
\im\chi^{\text{irr}}_c(\bq,\omega)=\frac{q^2}{e^2\omega}\re\sigma_{||}(\bq,\omega)\label{17}
\eea

Now we are going to invoke the result by Mishchenko, Reizer, and Glazman,\cite{mishchenko:2004} who showed that  the $T=0$ longitudinal conductivity of a 2D electron system is given by   
\bea
\re\sigma_{||}(\bq,\omega)=\frac{e^2}{12\pi^2} \frac{q^2}{k_{\text{F}} ^2}\ln\frac{\vf\kappa}{|\omega|}.
\label{sigma}
\eea
This result applies to the range of frequencies of interest to us, i.e., $\vf q\ll \omega\ll \vf\kappa$.
A factor of $q^2/k_{\text{F}}^2$ in Eq.~(\ref{sigma}) reflects the fact that, since the charge current is conserved in a Galilean-invariant FL, the dissipative part of its conductivity  must vanish at $q=0$. It is this factor that suppresses the $q^2/\omega$ contributions from diagrams 1-5 in Fig.~\ref{fig1}. Combining Eqs.~(\ref{17}) and (\ref{sigma}), we find
\bea
\im\chi^{\text{irr}}_c(\bq,\omega)=\frac{1}{12\pi^2 \vf k_{\text{F}} ^2}\frac{q^4}{\omega}\ln\frac{\vf\kappa}{|\omega|}.
\eea
Therefore, the leading term in $\im\chi^{\text{irr}}_c(\bq,\omega)$ of a Galilean-invariant system
scales as $q^4/\omega$ (modulo a logarithmic factor).  

A 3D analog of Eq.~(\ref{sigma}) for  the longitudinal conductivity is not available. However, it is known that the plasmon damping coefficient, $\gamma(q)$, scales as $q^2$ in 3D.\cite{dubois:1969}
[$\gamma(q)$ is defined such that the plasmon pole is located at $\omega=\omega_p(q)-i\gamma(q)$, where $\omega_p(q)$ is the plasma frequency in the absence of damping.]
From Eq.~(\ref{fullcharge}), it is easy to deduce that $\gamma=2\pi\re\sigma\left[\bq,\omega=\omega_p(q)\right]$.
Therefore, Re$\sigma\left[\bq,\omega=\omega_p(q)\right]\propto q^2$ in 3D as well which, according to Eq.~(\ref{17}), implies that $\im\chi^{\text{irr}}_c\left[\bq,\omega_p(q)\right]\propto q^4$.  It would be natural to expect that the frequency dependence is also $1/\omega$ (up to a logarithmic factor).
 We thus surmise that $\im\chi^{\text{irr}}$ in 3D scales with $\omega$ and $q$ in the same way as in 2D, i.e., 
\beq
\im\chi^{\text{irr}}_c(\bq,\omega)\propto \frac{q^4}{\omega}.
\label{chic3D}
\eeq


Note that  diagrams for the irreducible part of the charge susceptibility cancel each other also in the one-dimensional (1D) case for a linearized spectrum, e.g., for $\epsilon^{\pm}_\bk=\pm \vf (k\mp k_{\text{F}} )$ (Ref.~\onlinecite{zyuzin:2014}).
Keeping  a curvature term ($\sim k^2/2m$) in the dispersion, one gets non-zero $\im\chi_c$  outside of the continuum,\cite{pustilnik:2006, imambekov:2012, teber:2006} but it is smaller than the corresponding result for the spin susceptibility also by a factor of $q^2/k_{F}^2\ll 1$. In the 1D case, the difference between the charge and spin channels receives a natural explanation within the bosonization technique, in
which  the charge channel is mapped onto free bosons while the spin channel is mapped onto the sine-Gordon model.
Although the cosine term in the sine-Gordon model is marginally irrelevant for the repulsive interaction between fermions, it does lead to damping of spin bosons and hence to a non-zero $\im\chi_s$ already for a linearized dispersion. To obtain damping of charge bosons, one needs to go beyond the Luttinger-liquid paradigm by retaining a finite fermionic mass.\cite{pustilnik:2006, imambekov:2012, teber:2006} 

The results of this section show, however, that the suppression of damping in the charge channel as compared to the spin one occurs in all dimensions and thus does not rely on such specifically 1D features, as integrability and spin-charge separation. The underlying mechanism is Galilean invariance, which suppresses the longitudinal conductivity, and thus the charge susceptibility, by a factor of $q^2/k_{\text{F}} ^2$. In contrast, the spin susceptibility is free of such a constraint. If Galilean invariance is broken by, e.g., lattice or spin-orbit interaction, one should expect damping  in the charge and spin channels to be comparable, i.e., $\im\chi_{c}^{\text{irr}}(\bq,\omega)$ should also scale as $q^2/\omega$.
\subsubsection{Full charge susceptibility}
Having analyzed the scaling form of the irreducible charge susceptibility in the previous section, we can now describe
the full charge susceptibility given by Eq.~(\ref{fullcharge}). The difference between the two susceptibilities is mainly due to the plasmon pole which is absent in the irreducible susceptibility but present in the full one. Taking the imaginary part of Eq.~(\ref{fullcharge}), 
we obtain
\bea
\im\chi_{c}(\bq,\omega)=\frac{q^2}{e^2\omega}\frac{\sigma'}{\left(1-A_D\frac{2\pi\sigma''}{\omega}\right)^2+\left(A_D\frac{2\pi\sigma
'}{\omega}\right)^2},\nn\\
\label{imchic}
\eea
where we abbreviated $\sigma'\equiv \re\sigma_{||}(\bq,\omega)$ and $\sigma''\equiv \im\sigma_{||}(\bq,\omega)$. To lowest order in the electron-electron interaction, $\sigma''=ne^2/m\omega$. Then $A_D\sigma''/\omega$ can be re-written as 
$\omega_p^2(q)/\omega^2$, where the plasmon frequency $\omega_p(q)\propto \sqrt{q}$ in 2D and $\omega_p(q)=\text{const}$ in 3D, upon which Eq.~(\ref{imchic}) acquires 
a more transparent form
\bea
\im\chi_{c}(\bq,\omega)=\frac{q^2}{e^2\omega}\frac{\sigma'}{\left[1-\frac{\omega^2_p(q)}{\omega^2}\right]^2+\left(A_D\frac{2\pi\sigma
'}{\omega}\right)^2},
\label{imchic2}
\eea
in which $\sigma'$ is obviously related to plasmon damping. 
Away from the immediate vicinity of the plasmon pole, the damping term can be neglected
and Eq.~(\ref{imchic2}) is further simplified to
\bea
\im\chi_{c}(\bq,\omega)=\frac{q^2}{e^2\omega}\frac{\sigma'}{\left[1-\frac{\omega^2_p(q)}{\omega^2}\right]^2}.
\label{imchic3}
\eea
For $\vf q\ll \omega\ll\omega_p(q)$, 
\bea
\im\chi_{c}(\bq,\omega)\approx \frac{q^2\omega^3}{e^2\omega_p^4(q)}\sigma'\propto\left\{
\begin{array}{ccc}
\omega^3\sigma',\\
q^2\omega^3\sigma',
\end{array}
\right.
\eea
where the first (second) line refers to the 2D (3D) case.
For a Galilean-invariant FL, $\sigma\propto q^2$ (up to a logarithmic factor in 2D),
and 
\bea
\im\chi_{c}(\bq,\omega)\propto\left\{
\begin{array}{ccc}
q^2\omega^3,\\
q^4\omega^3.
\end{array}
\right.
\eea
For $\omega\gg\omega_p(q)$, the denominator in Eq.~(\ref{imchic3}) can be replaced by unity
and the difference between the full and irreducible charge susceptibilities disappears. In this region, therefore,
\bea
\im\chi_{c}(\bq,\omega)\approx \im\chi^{\text{irr}}_{c}(\bq,\omega)\propto \frac{q^2}{\omega}\sigma'\propto \frac{q^4}{\omega},
\eea
where we again assumed the Galilean-invariant case at the last step (and omitted the logarithmic factor in 2D).
\begin{figure}[t]
\vspace{0.1in}
\includegraphics[width=1.0 \linewidth]
{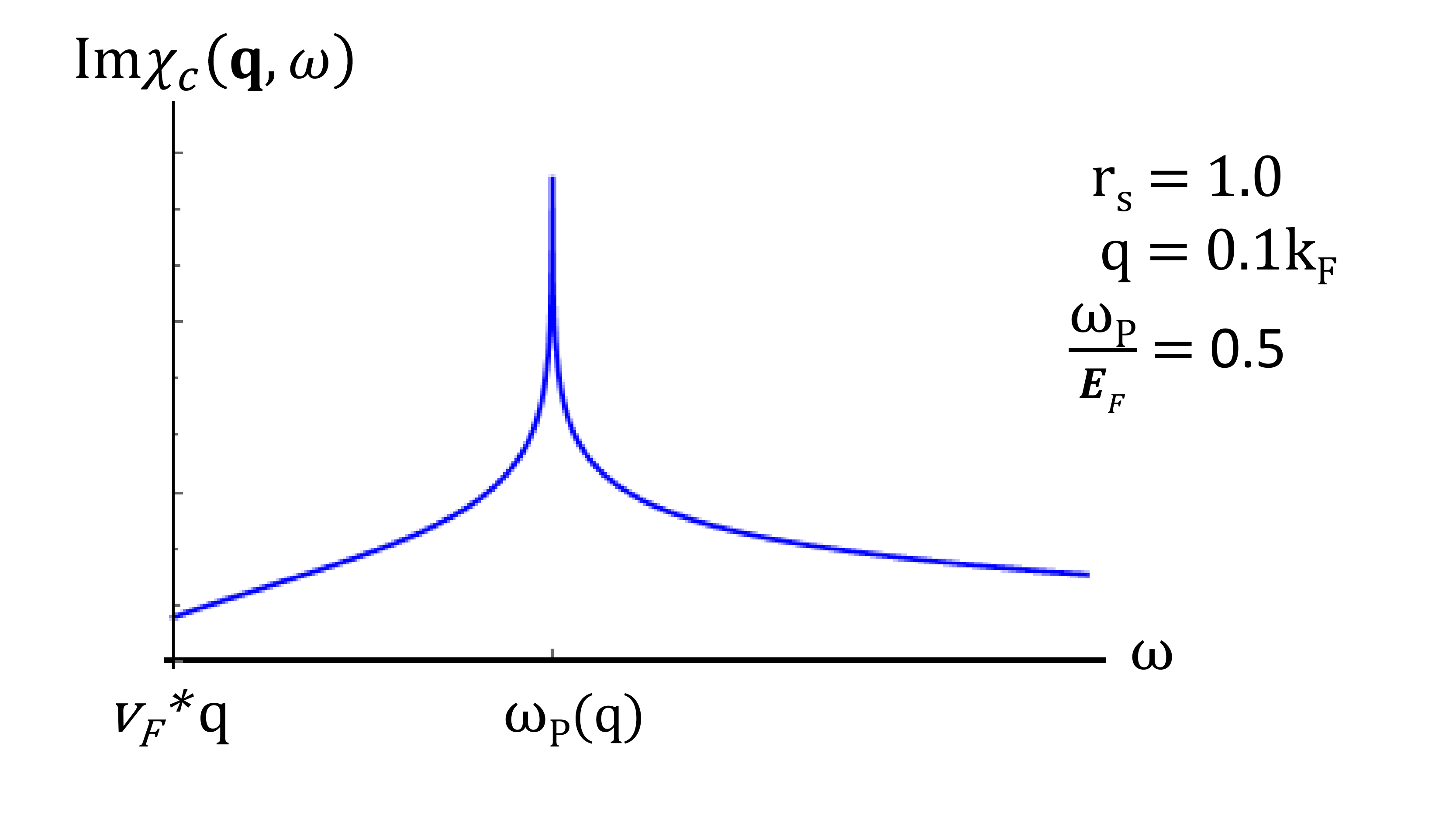}
\caption{Semi-log scale: Imaginary part of the charge susceptibility in 2D, as given by Eq.~(\ref{imchic2}). 
\label{fig:charge}}
\end{figure}

The imaginary part of the full charge susceptibility as a function of frequency is sketched  in Fig.~\ref{fig:spin-charge} by the dashed line. Figure \ref{fig:charge} shows (on a semi-logarthmic plot) the actual behavior of $\im\chi_{c}(\bq,\omega)$ in 2D, as given by Eq.~(\ref{imchic2}) for parameters specified in the legend. Note that the plasmon peak is flanked by two regions on the left and on the right, with asymptotic forms distinct both from the peak itself and from each other.

\subsection{Nematic susceptibility}
\label{sec:nematic}
In the previous sections, we considered the spin and charge susceptibilities which are related to the correlation functions of conserved quantities (spin and charge). Consequently, the spin and charge susceptibilities vanish at $q=0$ and finite $\omega$, which does affect their behavior at finite but small $q$: indeed, we found that $\im \chi_s(\bq,\omega)\propto q^2/\omega$  and $\im \chi_c(\bq,\omega)\propto q^4/\omega$ for a Galilean-invariant FL. On the other hand, a susceptibility related to a non-conserved quantity  does not have to vanish at $q=0$, and its behavior can be expected to differ significantly from the charge and spin cases considered earlier in this paper. The dynamic susceptibility of a nematic order parameter with a $d$-wave symmetry near a quantum critical point has recently been analyzed by Klein et al.\cite{klein:2018}  
For completeness,  we consider the spin susceptibility in the nematic channel with a $p$-wave symmetry, $\chi_{sc}(\bq,\omega)$, in a wide frequency interval, including 
the range of $\omega\gg \Ef$. 

Alternatively, $\chi_{sc}(\bq,\omega)$ can be viewed as the spin-current--spin-current correlation function, where the spin-current is defined as $J^{ij}_{s}(\bq)=\sum_\bk c^\dagger_{\bk+\bq/2} v^i_{\bk}\sigma^j c_{\bk-\bq/2}$. To be specific, we pick the $xz$ component of $J_s$. Due to in-plane rotational invariance, the result can be represent  as the half-sum of the $xz$ and $yz$ components of the susceptibility. Because the AL diagrams vanish due to spin traces, the leading-order correction to $\chi_{sc}(\bq,\omega)$ is given by diagrams 1-3 in Fig.~\ref{fig1}, where now the wiggle at the vertex denotes ${\bf v}\sigma^z$. Since $\chi_{sc}(\bq,\omega)$ is finite at $q=0$, we will set $q=0$ from the outset and study the frequency dependence of 
$\chi_{sc}(\omega)\equiv \chi_{sc}({\bf 0},\omega)$. Formally, the problem is equivalent to finding the current-current correlation function without the AL diagrams. This problem was considered in Ref.~\onlinecite{chubukov:2017}, where it was shown that the sum of diagrams 1-3 at $q=0$ can be written compactly as
\bwt
\bea
\label{sc}
\delta\chi_{sc}(\omega_m)=\int\int\int\int\frac{d^DQd^Dkd\Omega_ld\e_m}{(2\pi)^{2(D+1)}}&& \frac{\left(\bv_{\bk+\bQ}-\bv_\bk\right)^2U^{\text{dyn}}(\bQ,\Omega_l)}{\left[i(\Omega_l+\omega_m)-\epsilon_{\bk+\bQ}+\epsilon_\bk\right]\left[i\Omega_l-\epsilon_{\bk+\bQ}+\epsilon_\bk\right]}
G(\bk,\e_m)G(\bk+\bQ,\e_m+\Omega_l).
\nn\\
\eea
\ewt
Following an analogy with the current-current correlation function, $\chi_{j}(\omega)$, the result for $\delta\chi_{sc}(\omega)$ can be deduced without any computations. Indeed, the conductivity is related to $\chi_{j}(\omega)$ via $\re\sigma(\omega)=\im \chi_{j}(\omega)/\omega$. In its turn, the optical conductivity of a FL is of the Drude form: $\re\sigma(\omega)\propto 1/\omega^2\tau(\omega)$ with $1/\tau(\omega)\propto \omega^2$ for $\omega\ll \Ef$. Therefore, $\re\sigma(\omega)=\text{const}$ (this is the so-called ``FL foot''; see, e.g., Ref.~\onlinecite{maslov:2017b} and references therein). Consequently, $\im\chi_{j}(\omega)\propto\omega$ and, because $\chi_{sc}=\chi_{j}$ up to a factor of $e^2$, $\im\chi_{sc}(\omega)\propto \omega$ as well. The same argument applied to the $z=3$ quantum critical point, where $\sigma(\omega)\propto \omega^{-2/3}$ (Refs.~\onlinecite{kim:1994,eberlein:2016,chubukov:2017}), yields $\im\chi_{sc}(\omega)\propto \omega^{1/3}$ in agreement with the results of Ref.~\onlinecite{klein:2018}.

In line with the argument presented above, an explicit calculation for a FL with dynamically screened Coulomb potential gives (see Appendix \ref{sec:sc_low})
\bea
\label{sc_low}
\im\chi_{sc}(\omega)=\lambda_D
\left\{
\begin{array}{ccc}
\ln\left(\frac{\kf}{\kappa}\right)\omega,\\
\frac 32 \kf\omega,
\end{array}
\right.
\eea
for $\omega\ll \vf\kappa$, were $\lambda_D$ are again given by Eqs.~(\ref{lambda2}) and (\ref{lambda3}). The low-frequency scaling form, $\im\chi_{sc}(\omega)\propto \omega$, is expected to be valid for any FL, the only difference between the results of particular models being in the prefactor of the $\omega$ dependence. 

In contrast to the case of a conserved quantity, the susceptibility of a non-conserved quantity {\em increases} with frequency  for $\omega\gg \vf q$. However,  $\im\chi_{sc}(\omega)$ must decrease with $\omega$ at high enough frequencies, because electrons are not be able to follow very rapid oscillations of the external field. To see how this increase is curbed off,
we consider higher frequencies. Delegating the computational details to Appendices \ref{app:sc_interm} and \ref{app:sc_high}, we present here only the final results.

\begin{figure}[t]
\vspace{0.2in}
\includegraphics[width=1.0\linewidth]{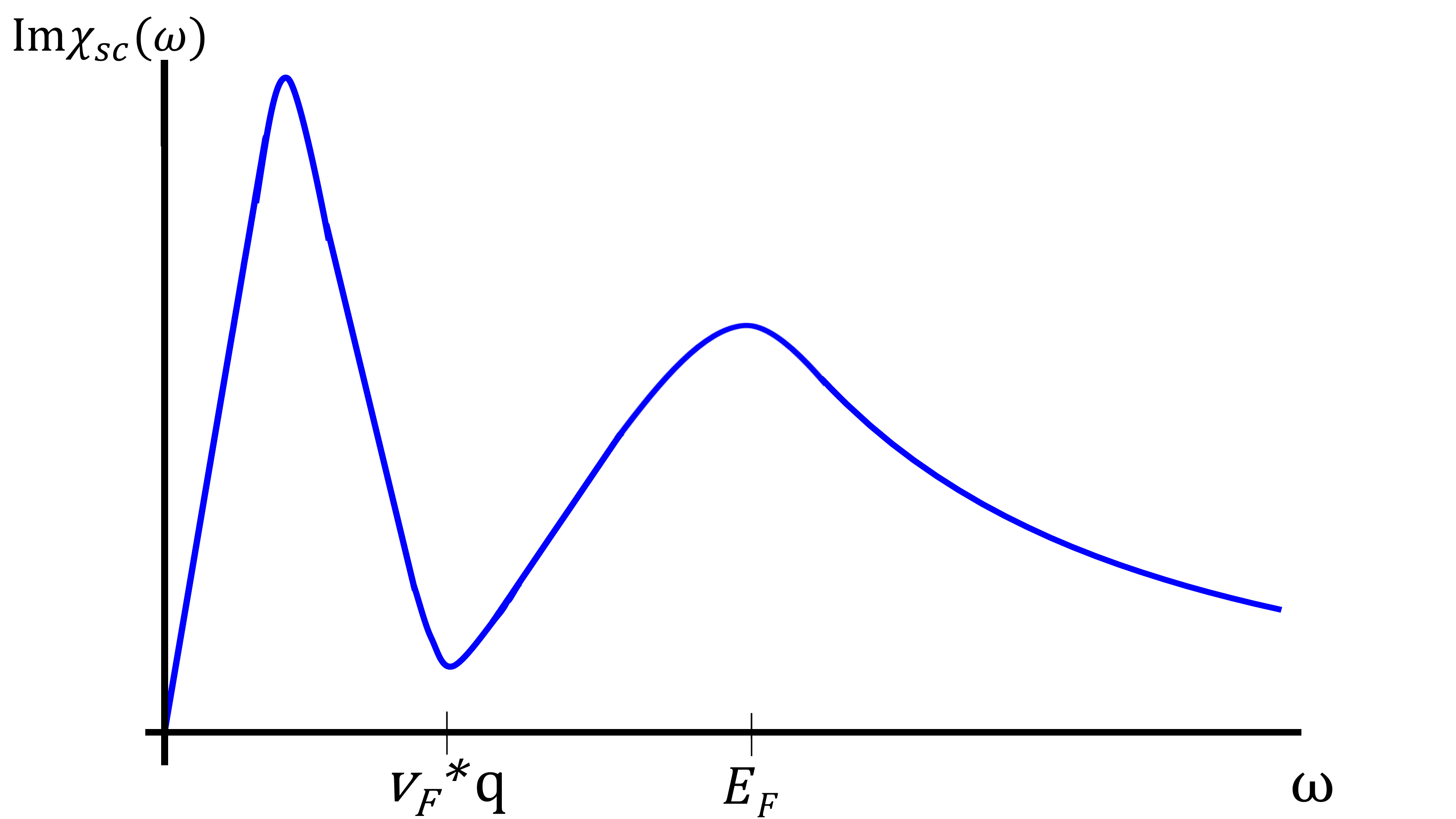}
\caption{A sketch of the imaginary part of the spin-current (nematic) susceptibility. For $\omega\gg\vf q$, the behavior is described by Eqs.~(\ref{sc_low}-
\ref{sc3di}).\label{fig:nem}}
\end{figure}

 In 2D,  the growth of $\im\chi_{sc}(\omega)$ with $\omega$ continues through the range $\vf\kappa\ll\omega\ll\Ef$, where
 \bea
 \label{sc2di}
 \im\chi_{sc}(\omega)=\lambda_2\omega\ln\frac{\Ef}{|\omega|},
 \eea
until $\im\chi_{sc}(\omega)$ reaches a maximum at $\omega=\Ef$. At even higher frequencies, $\omega\gg\Ef$, $\im\chi_{sc}(\omega)$ falls off as $1/\omega^2$:
 \bea
 \label{sc_high}
\im\chi_{sc}(\omega)=\frac{e^4}{\vf^2}\frac{E_{\text{F}} ^3}{\omega|\omega|}.
\eea
In 3D, $\im\chi_{sc}(\omega)$ is independent of $\omega$ for $\vf\kappa\ll\omega\ll E_F$:
\bea
\label{sc3di}
\im\chi_{sc}(\omega)=\frac{8\ln 2}{\pi^3}\frac{e^4}{\vf^2} \kf\Ef=\text{const}.
\eea
A decrease of $\im\chi_{sc}(\omega)$ with $\omega$ starts again at $\omega\sim\Ef$. One can show that $\im\chi_{sc}(\omega)\propto\text{sgn}\omega/|\omega|^{5/2}$ for  $\omega\gg \Ef$.
A sketch of  $\im\chi_{sc}(\omega)$ in a wide frequency interval is shown in Fig.~\ref{fig:nem}.

\section{Conclusions}
\label{sec:concl}
In this paper we have studied 
 the dynamical response of a FL  in the spin, charge, and nematic channels. First, we considered non-interacting quasiparticles, which technically amounts either to solving the FL kinetic equation without the collision integral or to resumming RPA series without taking corrections to the irreducible susceptibility into account. We solved the FL kinetic equation both analytically, for the Landau function containing up to first three harmonics, and numerically, for the Landau function corresponding to the statically screened Coulomb interaction. We showed that although the imaginary part of the susceptibility does exhibit an RPA-type singularity just below the particle-hole continuum boundary, i.e.,  $\im\chi(\bq,\omega)\propto \sqrt{\vf^*q-\omega}$ and $\im\chi(\bq,\omega)\propto 1/\ln^{2}(\vf^*q-\omega)$ for $\omega\to \vf^*q-0$  in 2D and 3D, respectively, this behavior becomes confined to a progressively narrower region near the boundary, as the number of harmonics in the Landau function is increased. For example, the square-root singularity for the screened Coulomb potential is visible only in the region $(\vf^*q-\omega)/\vf^* q\sim 10^{-6}$.

Next, we took into account the
 residual interaction between quasiparticles within a model of dynamically screened Coulomb potential. The main effect of such interaction is to produce a non-zero spectral weight of charge and spin fluctuations outside the particle-hole continuum. We showed that, at $T=0$ and in the absence of disorder, $\im\chi_s(\bq,\omega)$ behaves  as $q^2/\omega$ for $\omega\gg \vf^*q$ both in 2D and 3D. The behavior of the charge susceptibility depends strongly on whether the system is Galilean-invariant or not. If it is (which was the case considered in this paper),  the tail of $\im\chi_c(\bq,\omega)$ is suppressed by factor $(q/k_F)^2\ll 1$ as 
 compared to that of $\im\chi_s(\bq,\omega)$, i.e., $\im\chi_c(\bq,\omega)\propto q^4/\omega$. On a technical level, the suppression occurs as a result of a 
 partial 
 cancellation between the one-loop self-energy, ladder, and Aslamazov-Larkin diagrams. However,  the suppression receives a more natural explanation within the relation between the charge susceptibility and longitudinal optical conductivity, $\sigma'_{||}(\bq,\omega)$. In a Galilean-invariant system, $\sigma'_{||}(\bq,\omega)\propto q^2$, hence an extra factor of $q^2$ in $\im\chi_c(\bq,\omega)$. On the other hand, one should expect the charge and spin susceptibilities to be comparable if Galilean invariance is broken by, e.g., lattice or spin-orbit interaction. 

Although we obtained the $q^2/\omega$  asymptotic form of the spin  susceptibility in one-loop perturbation theory for the dynamically screened Coulomb potential, we believe that this form pertains to a FL of any kind. Indeed, a degenerate system of electrons exhibits a FL behavior for $\omega$ much smaller that effective plasma frequency, $\vf \kappa$ (Ref.~\onlinecite{agd:1963}), which is where our asymptotic form is valid.

We also found the asymptotic forms of the spin, charge, and nematic susceptibilities above the ultraviolet scales of the problem, i.e., the effective plasma frequency and Fermi energy. In contrast to the $q^2/\omega$ form, however, these forms are specific for the Coulomb system. 

We hope that the results of this paper will be useful in discriminating between the FL and non-FL behaviors observed in the experiments on nominally non-FL materials, such as copper-oxide superconductors.\cite{mitrano:2017}

\acknowledgements
We are grateful to A. V. Chubukov, S. Maiti, E. G. Mishchenko, and P. Kumar for stimulating discussions. D.L.M. acknowledges financial support from National Science Foundation under grant DMR-1720816. 
Part of his work was performed at the Aspen Center of Physics, supported by National Science Foundation under grant NSF PHY-160761, and Kavli Institute for Theoretical Physics,  supported by the National Science Foundation under grant NSF PHY-1748958.

\begin{widetext}
\appendix
\section{Diagrams for spin, charge, and nematic susceptibilities.}
\label{sec:appb}
In this Appendix, we present the calculations of the irreducible susceptibility (polarization bubble) beyond the RPA level.  The self-energy and ladder diagrams for the spin and nematic susceptibilities are discussed in Secs.~\ref{sec:se+l_2D} and \ref{sec:spin-current}, respectively. In Sec.~\ref{sec:AL}, we demonstrate that the AL diagrams cancel the self-energy and ladder ones in the charge channel.
\subsection{Self-energy and ladder diagrams for the spin susceptibility}
\label{sec:se+l_2D}
\subsubsection{Combining the diagrams}
In this section, we calculate the sum of diagrams 1 and 2 (self-energy) and 3 (ladder) in Fig.~\ref{fig1}.
The sum of the 
 two self-energy diagrams is given by
\begin{align}
\delta\chi_{\text{SE}}(q)=\delta\chi^{(1)}(q)+\delta\chi^{(2)}(q)=
-2 \int_{k} G_{k}G_{k+q} \left( G_{k+q} \Sigma_{k+q}+ G_{k}\Sigma_{k} \right),
\end{align}
where $\Sigma_{k}$ is the self-energy
\begin{align}
\Sigma_{k} = - \int_{Q} U_Q G_{k+Q},\label{a2}
\end{align}
\begin{align}
U_Q = \frac{1}{U^{-1}_0({\bf Q}) - \Pi^{(0)}_Q}
\end{align}
is the dynamically screened Coulomb interaction, and $U_0(\bQ)$ is the bare Coulomb potential.
Whenever it does not lead to confusion, will be using the notation $k = (\bk,\e_n)$,  $Q = (\bQ,\Omega_l)$  and $q=(\bq,\omega_m)$. Also, $\int_k$ is a short-hand for
$\int d^Dk/(2\pi)^D\int d\e_n/(2\pi)$, etc. and $G_k$ is a short-hand for $G(\bk,\e_n)$, etc.

Using the identity
\bea
G_{k+q} G_k=\frac{1}{i\omega_m-\epsilon_{\bk+\bq}+\epsilon_\bk}\left(G_k-G_{k+q}\right),\label{a6}
\eea
we represent $\delta\chi_{\mathrm{SE}} $ as the sum of two parts:  $\delta\chi_{\mathrm{SE}}(q)= \delta\chi^{(I)}_{\mathrm{SE}}(q)+ \delta\chi^{(II)}_{\mathrm{SE}}(q)$, where 
\begin{align}
\delta\chi^{(I)}_{\mathrm{SE}}(q)= 
2
\int_k \frac{G_k G_{k+q}\left( \Sigma_{k+q}-\Sigma_k  \right)}{(i \omega_m - \epsilon_{\bk+\bq}+\epsilon_\bk)}
\end{align}
and
\begin{align}
\delta\chi^{(II)}_{\mathrm{SE}}(q)= 
2
\int_k \frac{G^2_k \Sigma_k-G^2_{k+q} \Sigma_{k+q}}{(i \omega_m - \epsilon_{\bk+\bq}+\epsilon_\bk)} = 2\int_k G^2_k \Sigma_k \left[ \frac{1}{i \omega_m - \epsilon_{\bk+\bq}+ \epsilon_\bk} - \frac{1}{i \omega_m - \epsilon_\bk + \epsilon_{\bk-\bq}}\right].\label{b8}
\end{align}
For $|\bq|\ll k_{\text{F}} $, the fractions in the square brackets in Eq.~(\ref{b8}) can be expanded as
\begin{align}
\delta\chi^{(II)}_{\mathrm{SE}}(q)= 
2\frac{\bq^2}{m}\int_k G^2_k \Sigma_k  \frac{1}{\left(i \omega_m - \vf\hat\bk\cdot\bq\right)^2}.\label{a8}
\end{align}
Next, we integrate the combination $G^2_k\Sigma_k$ over $\e_n$ and $\epsilon_\bk$, assuming that relevant $|\bQ|$ are small:  $|\bQ|\ll k_{\text{F}} $. This yields 
\begin{align}
\int d{\epsilon_\bk}\int \frac{d\e_n}{2\pi} G^2_k \Sigma_k = - \int \frac{d^DQ}{(2\pi)^D} \frac{d\Omega_l}{2\pi}\int{d\epsilon_\bk}\int\frac{d\e_n}{2\pi} G^2_k G_{k+Q} U_Q =\int \frac{d^DQ}{(2\pi)^D}\int \frac{d\Omega_l}{2\pi}U_Q \frac{\vf\hat\bk \cdot \bQ}{(i \Omega_l - \vf\hat\bk \cdot \bQ)^2}.
\end{align}
The fraction in the integrand of the last equation above is odd upon a simultaneous change of variables $\Omega_l \rightarrow  -\Omega_l$ and $\bQ \rightarrow - \bQ$, while $U_Q$ is even under either of these two operations.  Therefore, $\delta\chi^{(2)}_{\mathrm{SE}}(q) = 0$. The assumption of $|\bQ|\ll k_{\text{F}} $ is justified because the range if integration over $|\bQ|$ is effectively limited by the (inverse) screening radius, $\kappa$, which must be chosen to be smaller than $k_{\text{F}} $ for the perturbation theory to be under control. Keeping higher order of $|\bf Q|$ would result in $\kappa/k_{\text{F}} $ corrections. Note that the vanishing of $\delta\chi^{(2)}_{\mathrm{SE}}(q)$ occurs  regardless of the choice of $\omega$. This circumstance will be used later for finding the high-frequency tail of the susceptibility.

Using Eq.~(\ref{a2}) for $\Sigma_k$ and applying identity (\ref{a6}) again, we re-write $\delta\chi^{(I)}_{\mathrm{SE}}$ as
\begin{align}
\delta\chi^{(I)}_{\mathrm{SE}}(q) = -2
 \int_{k,Q} \frac{(G_k- G_{k+q)}) (G_{k+Q}- G_{k+Q+q})}{(i \omega_m - \epsilon_{\bk+\bq}+ \epsilon_\bk) ^2} U_Q.
\end{align}
With the help of Eq.~(\ref{a6}), the ladder diagram (diagram 3 in Fig.~\ref{fig1}) can be re-written as
\begin{align}
 \delta\chi^{(3)}(q) =
 2
  \int_{k,Q} \frac{(G_k- G_{k+q)}) (G_{k+Q}- G_{k+Q+q})}{(i \omega_m - \epsilon_{\bk+\bq}+ \epsilon_\bk) (i \omega_m - \epsilon_{\bk+\bq+\bQ} + \epsilon_{\bk+\bQ})} U_Q.
\end{align}
For the sum of the self-energy and ladder diagrams we then find \begin{align}
\delta\chi_s(q)= \delta\chi^{(I)}_{\mathrm{SE}}(q) +  \delta\chi^{(3)}(q) = 
-2
\int_{k,Q} \frac{(G_k- G_{k+q)}) (G_{k+Q}- G_{k+Q+q})}{i \omega_m - \epsilon_{\bk+\bq}+ \epsilon_\bk } U_Q\left[\frac{1}{i \omega_m - \epsilon_{\bk+\bq}+ \epsilon_\bk} -  \frac{1}{i \omega_m - \epsilon_{\bk+\bq+\bQ} + \epsilon_{\bk+\bQ}}\right].
\end{align}
It will prove to be convenient to re-write the equation above in a symmetric form by relabeling $Q= p-k$:
\bea
\delta\chi_s(q)&=&
-2
\int_{k,p} \frac{(G_k- G_{k+q)}) (G_{p}- G_{p+q})}{i \omega_m - \epsilon_{\bk+\bq}+ \epsilon_\bk} U_{p-k}\left[ \frac{1}{i \omega_m - \epsilon_{\bk+\bq}+ \epsilon_\bk} -\frac{1}{i \omega_m - \epsilon_{\bp+\bq} + \epsilon_\bp} \right]\nn\\
&=&
-2
\int_{k,p} (G_k- G_{k+q)}) (G_{p}- G_{p+q})U_{p-k}\frac{
\epsilon_{\bk+\bq}-\epsilon_\bk-\epsilon_{\bp+\bq}+\epsilon_\bp}
{(i \omega_m - \epsilon_{\bk+\bq}+ \epsilon_\bk)^2 (i \omega_m - \epsilon_{\bp+\bq} + \epsilon_\bp)}.
\eea
We now symmetrize the equation above by re-writing $\delta\chi_s(q)=(1/2)\delta\chi_s(q)+(1/2)\delta\chi_s(q)$ and interchanging $k \leftrightarrow p$ in one of the two terms, while keeping in mind $U_{Q}$ is an even function of $Q$.  This gives\cite{geldart:1970,maslov:2017}  
\begin{align}
\delta\chi_s(q)=-\int_{k,p} (G_k- G_{k+q)}) (G_{p}- G_{p+q})U_{p-k}\frac{
\left(\epsilon_{\bk+\bq}-\epsilon_\bk-\epsilon_{\bp+\bq}+\epsilon_\bp\right)^2}{(i \omega_m - \epsilon_{\bp+\bq} + \epsilon_\bp)^2 (i \omega_m - \epsilon_{\bk+\bq}+ \epsilon_\bk)^2}.
\end{align}
 Relabeling back  $p=k+Q$, we arrive at following form
 \begin{align}
\delta\chi_s(q)=-\int_{k,Q} (G_k- G_{k+q)}) (G_{k+Q}- G_{k+Q+q}) U_{Q}\frac{
\left(\epsilon_{\bk+\bq}-\epsilon_\bk-\epsilon_{\bk+\bQ+\bq}+\epsilon_{\bk+\bQ}\right)^2}{(i \omega_m - \epsilon_{\bk+\bQ+\bq} + \epsilon_{\bk+\bQ})^2 (i \omega_m - \epsilon_{\bk+\bq}+ \epsilon_\bk)^2}. \label{A14}
\end{align}
Equation (\ref{A14}) is a general result valid for $|\bq|\ll k_{\text{F}} $ and arbitrary $\omega$. In what follows,  we will analyze the various limiting cases.

For a parabolic dispersion, $\epsilon_\bk=(k^2-k_{\text{F}} ^2)/2m$, Eq.~ (\ref{A14}) is reduced to
\begin{align}
\delta\chi_s(q)=-\frac{1}{m^2} \int_{k,Q} (G_k- G_{k+q)}) (G_{k+Q}- G_{k+Q+q})U_{Q}\frac{
\left(\bq\cdot\bQ\right)^2}{(i \omega_m - \epsilon_{\bk+\bQ+\bq} + \epsilon_{\bk+\bQ})^2 (i \omega_m - \epsilon_{\bk+\bq}+ \epsilon_\bk)^2}.\label{A15}
\end{align}
\subsubsection{Low frequencies: $\omega\ll \vf\kappa$}
In this section, we focus on the case of parabolic dispersion. First, we consider the region of $\omega$ small compared to the energy scale set by the interaction, i.e., $\omega\ll \vf\kappa$.  In this case, typical momentum transfers are on the order of the (inverse) screening radius, i.e., 
$|\bQ|\sim\kappa\ll k_{\text{F}} $. (To be more precise, the final integral over $|\bQ|$ in 2D will be shown to be logarithmic, with the base of support in the region $\kappa\ll |\bQ|\ll k_{\text{F}} $.) Therefore, the dispersions can be expanded as $\epsilon_{\bk+\bQ+\bq}- \epsilon_{\bk+\bQ}\approx \bv_{\bk+\bQ}\cdot\bq\approx \vf\hat\bk\cdot\bq$ and $\epsilon_{\bk+\bq}-\epsilon_\bk\approx \vf\hat\bk\cdot\bq$ as before. With these simplifications, Eq.~(\ref{A15}) is reduced to 
\bea
\delta\chi_s(q)=- \frac{1}{m^2}\int_{k,Q}   \left(G_k-G_{k+q}\right)\left(G_{k+Q}-G
_{k+q+Q}\right)U_Q \frac{(\bq\cdot \bQ)^2}{(i \omega_m - \vf\hat\bk\cdot\bq)^4} .
\eea
Next, we integrate the products of the Green's functions in the equation above first over $\e_n$ and then over $\epsilon_\bk$, and neglect $|\bq|$ compared to $|\bQ|$ in the final result. This yields
\bea
\label{A17}
\delta\chi_s(q)=- \frac{iN_F}{2m^2}\int \frac{d\hat\bk}{{\mathcal O_D}}\int_{Q} \left[ \frac{2\Omega_l}{i \Omega_l - \vf\hat\bk \cdot \bQ} - \frac{\Omega_l-\omega_m}{i (\Omega_l- \omega_m) - \vf\hat\bk\cdot\bQ }- \frac{\Omega_l+\omega_m}{i (\Omega_l + \omega_m) - \vf\hat\bk\cdot\bQ}\right] 
U_Q \frac{(\bq\cdot \bQ)^2}{(i \omega_m - \vf\hat\bk\cdot\bq)^4},\nn\\
\eea
where $N_F$ is the density of states at the Fermi energy per two spin orientations, $\int d\hat\bk$ stands for the integral over the direction of $\bk$, ${\mathcal O_2}=2\pi$, and ${\mathcal O_3}=4\pi$.

Now we perform a standard trick of decomposing the momentum transfer $\bQ$ into components perpendicular ($Q_\perp$) and tangential ($Q_{||}$) to the Fermi surface at point $\bk$, while assuming that $|Q_\perp|\ll |Q_{||}|$.

The rest of the calculations differs somewhat between the 2D and 3D cases because of the differences in the geometry. We follow the 2D case in detail, and then just give the result for the 3D one. 
\paragraph{2D.}
In 2D, 
we re-write the dot product  $\bq \cdot \bQ$ as \bea\bq \cdot \bQ = |\bq| |\bQ| \cos(\theta_{\bk\bq}+ \theta_{\bk\bQ}) =|\bq|(Q_{\perp}\cos\theta_{\bk\bq}- Q_{||} \sin\theta_{\bk\bq})\approx - |\bq|Q_{||} \sin\theta_{\bk\bq},
\eea
where $\theta_{{\bf a}{\bf b}}$ is the angle between vectors ${\bf a}$ and ${\bf b}$. 
At the last step we employed the condition $Q_{\perp}\ll Q_{||}$ (recall that  $Q_{||}$ is a scalar in 2D). 
Under the  same assumption, $\bQ$ in the interaction potential can be replaced by $Q_{||}$, and the integral over $Q_\perp$ can be carried out using
\bea
\int^\infty_{-\infty} \frac{dx}{iy-x}=-i\pi \text{sgn}y.
\eea
 The integral over $\theta_{\bk\bq}$ is given by
\bea 
\int_{0}^{2\pi} \frac{d\theta_{\bk\bq}}{2\pi}  \frac{ \sin^2\theta_{\bk\bq} }{(i \omega_m - \vf |\bq|\cos\theta_{\bk\bq})^4} = \frac{|\omega_m|}{2\left(\omega_m^2 + \vf^2 \bq^2\right)^{\frac{5}{2}}}.
\eea
After these two steps, $\delta\Pi(q)$ is reduced to
\bea
\delta\chi_s(q) = -\frac{N_F}{4
m^2\vf}\frac{\bq^2|\omega_m|}{\left(\omega_m^2 + \vf^2 \bq^2\right)^{\frac{5}{2}}}\int^\infty_0\frac{dQ_{||}}{2\pi}  \int^\infty_{-\infty}\frac{d\Omega_l}{2\pi}  \left(2|\Omega_l|-|\Omega_l-\omega_m|
-|\Omega_l+\omega_m|\right)Q_{||}^2U_{Q_{||},\Omega_l}.\nn \\
\eea

Now we will simplify the form of the interaction potential. 
As we said before, typical $|\bQ|\approx |Q_{||}|$ are expected to be on the order of $\kappa$, whereas typical energy transfers ($\Omega$) are expected to be on the order of the external frequency: $\Omega\sim \omega$. For external frequencies in the interval $\omega\ll \vf\kappa$, we can take the limit of $\Omega\ll \vf |\bQ|$, when the Matsubara form of the free-electron bubble (in 2D) can be approximated as 
\begin{align}
\Pi^{(0)}_Q
= -N_F \left[ 1 -\frac{\vert\Omega_l\vert}{\vf |Q_{||}| }  \right].
\end{align}
Accordingly, the dynamic part of the Coulomb interaction is reduced to\begin{align}
U^{\text{dyn}}_{Q_{||},\Omega_l}=
\frac{2\pi e^2 \kappa}{\left(|Q_{||}|
+ \kappa\right)^2}
\frac{\vert\Omega_l\vert}{\vf  |Q_{||}| }.\label{a5}
\end{align}
The integral over $\Omega_l$ yields
\bea
\int_{-\infty}^{\infty}d\Omega_l |\Omega_l| \left(2|\Omega_l|- |\Omega_l-\omega_m|- |\Omega_l+ \omega_m| \right) = -\frac 23|\omega_m|^3.
\eea
The final result is obtained by integrating over $Q_{||}$ to logarithmic accuracy:
\bea
\delta_s\chi(q) &=&  \frac{q^2 e^2 \kappa N_F}{
12 \pi \vf^2m^2} \frac{\omega_m^4 \bq^2}{\left(\omega_m^2+ \vf^2 \bq^2\right)^{\frac{5}{2}}} \int_{0}^{k_{\text{F}} } dQ_{||} \frac{Q_{||}}{(Q_{||} + \kappa)^2} =  \frac{ e^4}{
6 \pi^2 \vf^2} \frac{\omega_m^4 \bq^2}{\left(\omega_m^2+ \vf^2 \bq^2\right)^{\frac{5}{2}}} \ln\frac{k_{\text{F}} }{\kappa}.
\eea
At this step, we see that typical $Q_{||}$ are indeed in the interval $\kappa\lesssim Q_{||}\lesssim k_{\text{F}} $, and thus the condition $\Omega_l/\vf|Q_{||}|\ll 1$ is satisfied. Upon analytic continuation, we obtain the top line of Eq.~(\ref{chi_s_main}) in the main text.

\paragraph{3D.} In 3D, free-electron polarization bubble in the quasistatic limit is of the same form as in 2D 
up to a numerical prefactor
\begin{align}
\Pi^{(0)}_Q = -\frac{mk_{F}}{2\pi^2}\left(1 - \frac{\pi}{2}\frac{|\Omega_l|}{\vf |\bQ|}\right).
\end{align}
The rest of the calculation differs only in that there is an additional integral over the azimuthal angle but this does not really complicate the matters. Without repeating the same steps as in the 2D case, we simply quote the final result 
\begin{align}
\delta\chi_s(q) = \frac{mk_{F}}{
18 \pi^2} \frac{\bq^2 \vert\omega_m\vert^3}{\left( \omega_m^2 + \vf ^2\bq^2 \right)^2} \frac{e^4}{\vf \kappa},
\end{align}
where $\kappa^2=4e^2mk_{\text{F}} /\pi$ is the inverse screening radius in 3D. Upon analytic continuation, this gives the bottom line in Eq.~(\ref{chi_s_main}).
\subsubsection{Intermediate frequencies: $\vf\kappa\ll\omega\ll E_{\text{F}} $}
\label{app:s_interm}
\paragraph{2D.}
In 2D, the only change compared to the case considered in the previous section is that the logarithmic integral over $Q$ needs to be cut at $|\omega_m|$ rather than at $\vf\kappa$. Then the result valid at all frequencies below the Fermi energy can be written as 
\bea
\delta_s\chi(q) =  \frac{ e^4}{
6 \pi^2 \vf^2} \frac{\omega_m^4 \bq^2}{\left(\omega_m^2+ \vf^2 \bq^2\right)^{\frac{5}{2}}} \ln\frac{E_{\text{F}} }{\max\{\vf\kappa,|\omega_m|\}}.
\eea
Upon analytic continuation and for $\omega\gg \vf\kappa$ the last equation gives the result in  Eq.~(\ref{2di}) of the main text.

\paragraph{3D.}In 3D, the analysis is more involved. We go back to Eq.~(\ref{A17}), neglect the $\vf\hat\bk\cdot\bq$ term compared to $\omega_m$, subtract off the static potential, and integrate over $\hat\bk$. This yields
\bea
\label{3D1}
\delta\chi_s(q)=-\frac{N_F}{m^2\omega_m^4}\int_Q&&\frac{(\bq\cdot\bQ)^2}{\vf|\bQ|} U^{\text{dyn}}_Q\nn\\
&&\times\left[2\Omega_l\tan^{-1}\frac{\vf|\bQ|}{\Omega_l}-(\Omega_l-\omega_m)\tan^{-1}\frac{\vf|\bQ|}{(\Omega_l-\omega_m)}-(\Omega_l+\omega_m)\tan^{-1}\frac{\vf|\bQ|}{(\Omega_l+\omega_m)}\right],
\eea
where
\bea
U^{\text{dyn}}_{Q}=U^{\text{dyn}}(\bQ,\Omega_l)=U(\bQ,\Omega_l)-U(\bQ,0).
\eea
We assume first and verify later that typical integration variables are in the range
$\vf|\bQ|\gtrsim|\Omega_l|\sim |\omega_m|\gg \vf\kappa$. The first inequality sign ($\gtrsim$) is to be understood in the logarithmic sense, while the last one is guaranteed by our the choice of the external frequency. Under these conditions, the dynamic interaction can be approximated by
\bea
\label{Udi}
U^{\text{dyn}}_{Q}\approx \frac{1}{N_F}\left(\frac{\kappa}{|\bQ|}\right)^4\frac{\Omega_l}{\vf|\bQ|}\tan^{-1}\frac{\vf|\bQ|}{\Omega_l}.
\eea
Substituting this form into Eq.~(\ref{3D1}) and rescaling the variables as $x=\Omega_l/\omega_m$ (for $\omega_m>0$) and $y=\vf|\bQ|/\omega_m$, we obtain
\bea
\label{B32}
\delta\chi_s(q)=-\frac{1}{6\pi^2}\frac{\bq^2\kappa^4}{\vf m^2\omega_m^2}\int^{E_{\text{F}} /\omega_m}_0 \frac{dy}{y^2} f(y) 
\eea 
where
\bea
f(y)=\int^\infty_0 dx x \tan^{-1}\left(\frac{y}{x}\right)\left[2x\tan^{-1}\left(\frac{y}{x}\right)-(x-1)\tan^{-1}\left(\frac{y}{x-1}\right)-(x+1)\tan^{-1}\left(\frac{y}{x+1}\right)\right].\label{A33}
\eea
In Eq.~(\ref{B32}) we retained the upper limit of integration at $|\bQ|=k_{\text{F}} $ in anticipation of a logarithmic divergence. To analyze the behavior of $f(y)$ for $y\ll 1$, we notice that a formal expansion in $y$ leads to a singularity of the integrand at $x=1$. Therefore, the integral is controlled by a narrow region around $x=1$. Introducing a new variable $z=(x-1)/y$ and setting $z=0$ in all but the last term in the square brackets, we obtain
\bea
f(y\ll 1)=2y^3\int^\infty_0dz\left(1-z\tan^{-1}\frac{1}{z}\right)=\frac{\pi}{2}y^3.
\eea
Therefore, the $1/y^2$ singularity at $y\to 0$ in Eq.~(\ref{A32}) is canceled.  For $y\gg 1$, we apply the identity $\tan^{-1}x=\frac{\pi}{2}-\tan^{-1}\frac{1}{x}$ to the square bracket in Eq.~(\ref{A33}) and expand the resultant expression in $1/y$. This yields
\bea
f(y\gg 1)=2y\int^\infty_0 dx \frac{x}{x^2+y^2}\tan^{-1}\left(\frac{y}{x}\right)=(\pi\ln 2) y.
\label{A35}
\eea
Therefore,  the remaining integral over $y$ is indeed logarithmic and can be solved in the leading logarithmic approximation with the result
\bea
\delta\chi_s(q)=-\frac{\ln 2}{6\pi}\frac{\kappa^4}{m^2\vf}\frac{\bq^2}{\omega_m^2}\ln\frac{E_{\text{F}} }{|\omega_m|}.
\eea
A non-zero imaginary part of $\delta\chi_s$ comes from the analytic continuation of the logarithmic factor. After analytic continuation, we arrive at Eq.~(\ref{3di}) of the main text.
\subsubsection{High frequencies: $\omega\gg E_{\text{F}} $}
\label{app:spin_high}
 In the preceding section, we considered the case when the external frequency and momentum, $\omega$ and $|\bq|$, are small compared to $E_{\text{F}} $ and $k_{\text{F}} $, respectively. Now we focus on the high-frequency regime, where $\omega\gg E_{\text{F}} $, while $|\bq|$ is still small, and consider only the 2D case. We recall that Eq.~(\ref{A14}) was derived without any restrictions on $\omega$. Since the integrand is already proportional to $\bq^2$ while we assume that $\vf |\bq|\ll \omega$, we set $|\bq|=0$ in the rest of the expression and obtain
 \begin{align}
\delta\chi_s(q)
=-\frac{1}{m^2\omega_m^4} \int_{k,Q} (G_k- G_{k+q)}) (G_{k+Q}- G_{k+Q+q})\left(\bq\cdot\bQ\right)^2U^{\text{dyn}}_{Q},
\label{b14}
\end{align}
where $U^{\text{dyn}}_{Q}$ is the dynamic part of the interaction.
Integrating the products of the Green's functions in the equation above over $\e_m$ and $\bk$, we obtain a combination of polarization bubbles
\begin{align}
\delta\chi_s(q)
=-\frac{1}{m^2\omega_m^4} \int_{\bk,Q}n_F(\epsilon_{\bk}) 
\left(\bq\cdot\bQ\right)^2U_{Q}
\left[2\Pi(\bQ,\Omega_l)-\Pi(\bQ,\Omega_l+\omega_m)-\Pi(\bQ,\Omega_l-\omega_m)\right].
\label{b15}
\end{align}
We first assume and then verify that  in the current regime $\Omega_l\sim\omega$ and $\bQ^2/m\sim \omega$, i..e,
$|\bQ|\sim \sqrt{m\omega}\gg k_{\text{F}} $. 

  The polarization bubble for $|\bQ|\gg k_{\text{F}} $ can be approximated as\cite{kashuba:2002,suhas:2005,chubukov:2010} \bea
\Pi(\bQ,\Omega_l)&=&\int \frac{d^2k}{(2\pi)^2} n_F(\epsilon_\bk)\left[\frac{1}{i\Omega_l-\epsilon_{\bk+\bQ}+\epsilon_\bk}-\frac{1}{i\Omega_l-\epsilon_{\bk}+\epsilon_{\bk-\bQ}}\right]
\nn\\
&&\approx\int \frac{d^2k}{(2\pi)^2} n_F(\epsilon_\bk)\left[\frac{1}{i\Omega_l-\frac{\bQ^2}{2m}}-\frac{1}{i\Omega_l+\frac{\bQ^2}{2m}}\right]=-n_0\frac{\frac{\bQ^2}{2m}}{\Omega_l^2+\left(\frac{\bQ^2}{2m}\right)^2}.\label{Pi_high}
\eea
Next, using the condition of $|\bQ|\gg k_{\text{F}} $, we approximate the dynamic interaction as
\bea
U^{\text{dyn}}_Q
&=&
(2\pi e^2)^2\frac{\Pi(\bQ,\Omega_l)-\Pi(\bQ,0)}{\left[|\bQ|-2\pi e^2 \Pi(\bQ,\Omega_l)\right] \left[|\bQ|-2\pi e^2 \Pi(\bQ,0)\right]}
\approx \frac{(2\pi e^2)^2}{\bQ^2}\left[\Pi(\bQ,\Omega_l)-\Pi(\bQ,0)\right]\label{Ud}.
\eea

Substituting  the last result into Eq.~(\ref{Ud}), we find
\bea
U^{\text{dyn}}_Q
=(2\pi e^2)^2 2m n_0\frac{1}{\bQ^4}\frac{\Omega_l^2}{\Omega_l^2+\left(\frac{\bQ^2}{2m}\right)^2}\label{Ud2}.\eea
Now we substitute Eqs.~(\ref{Pi_high}) and (\ref{Ud2}) back into Eq.~(\ref{sc2}) and rescale the variables as 
$x=\Omega_l/\omega_m$ and $y=\bQ^2/2m\omega_m$ (assuming that $\omega_m>0$). This yields
\bea
\delta\chi_s(q)=-\frac{n_0^2e^4}{m}\frac{\bq^2}{\omega_m^4}\int^\infty_{E_{\text{F}} /\omega_m} dy\int^\infty_{0}&& dx \frac{x^2}{x^2+y^2}\left[\frac{1}{(x+1)^2+y^2}+\frac{1}{(x-1)^2+y^2}-\frac{2}{x^2+y^2}\right].
\eea
As before, we kept the lower limit at $|\bQ|\sim k_{\text{F}} $ in anticipation of a logarithmic singularity. The integral over $x$ in the equation above yields $\pi/\left[2y(4y^2+1)\right]$ and thus the remaining integral over $y$ indeed diverges logarithmically at the lower limit. Solving this integral  to logarithmic accuracy and noting that the result must be an even function of $\omega_m$, we obtain
\bea
\delta\chi_s(q)=-\frac{\pi n_0^2e^4}{2m}\frac{\bq^2}{\omega_m^4}\ln\frac{|\omega_m |}{E_{\text{F}} }.
\eea
Performing analytic continuation, expressing $n_0$ in terms of the Fermi energy via $n_0=mE_{\text{F}} /\pi$  and taking the imaginary part, we arrive at Eq.~(\ref{2dh}) of the main text.

\subsection{Self-energy and ladder diagrams for the nematic susceptibility}
\label{sec:spin-current}
In this Appendix, we provide details of the calculation for the spin susceptibility in the nematic channel with angular momentum equal to unity. The sum of  diagrams 1-3 in Fig.~\ref{fig1} at $|\bq|=0$ is given by Eq.~(\ref{sc}) of the main text.
\subsubsection{Low frequencies: $\omega\ll \vf\kappa $}
\label{sec:sc_low}
We consider a parabolic single-particle dispersion, when ${\bf v}_{\bk+\bQ}-{\bf v}_\bk=\bQ/m$.  Expanding the dispersions to linear order in the momentum transfer $\bQ$, we integrate the product of the Green's functions in Eq.~(\ref{sc}) first over $\e_m$ and then over $\epsilon_\bk$ to obtain
\bea
\delta\chi_{sc}(\omega_m)=\frac{2N_F}{m^2}\int\int\int \frac{d^D\bQ d\Omega_l}{(2\pi)^{D+1}}
\frac{d\hat\bk}{{\cal O}_D}&& \frac{\bQ^2U^{\text{dyn}}(\bQ,\Omega_l)\vf\hat\bk\cdot\bQ}{\left[i(\Omega_l+\omega_m)-\vf\hat\bk\cdot\bQ\right]\left[i\Omega_l-\vf\hat\bk\cdot\bQ\right]^2}.
\eea

\paragraph{2D.}  Integrating over the angle between $\bk$ and $\bQ$  and dropping an odd in $\Omega_l$ part of the result, we obtain:
\bea
\delta\chi_{sc}(\omega_m)=\frac{N_F}{(2\pi)^2 m^2\omega_m^2}\int^\infty_0 d|\bQ| |\bQ|^3\int^\infty_{-\infty} 
d\Omega U^{\text{dyn}}(\bQ,\Omega_l)\left[\frac{|\Omega_l|}{\sqrt{\Omega_l^2+\vf^2\bQ^2}}-\frac{|\Omega_l+\omega_m|}{\sqrt{(\Omega_l+\omega_m)^2+\vf^2\bQ^2}}\right].\nn\\
\label{sc_2d}
\eea
As before, we first assume and then verify that typical $|\bQ|$ and $\Omega_l$  satisfy $|\bQ|\gtrsim \kappa\gg \Omega_l/\vf\sim \omega_m/\vf$.  Then the factor in the square brackets in the equation above is reduced to $|\Omega_l|-|\Omega_l+\omega_m|$, while the dynamic interaction can be replaced by
\bea
U^{\text{dyn}}(\bQ,\Omega_l)=N_F^{-1}\frac{\kappa^2}{(|\bQ|+\kappa)^2} \frac{|\Omega_l|}{\vf|\bQ|}.
\eea
The integral over $\Omega_l$ gives
\bea
\int^\infty_{-\infty} d\Omega_l |\Omega_l|\left(|\Omega_l|-|\Omega_l+\omega_m|\right)=-\frac 13 |\omega_m|^3.
\eea
From here, we already see that $\delta\chi_{sc}(\omega_m)\propto |\omega_m|$. Calculating the remaining integral over $|\bQ|$ to logarithmic accuracy, we find
\bea
\delta\chi_{sc}(\omega_m)=-\frac{1}{6\pi^2}\frac{e^4}{\vf^2} \ln\frac{k_{\text{F}} }{\kappa} |\omega_m|.
\eea
Upon analytic continuation, this gives the top line in Eq.~(\ref{sc_low}) of the main text.

\paragraph{3D.}
In 3D, angular integration yields
\bea
\delta\chi_{sc}(\omega_m)=\frac{N_F}{8\pi^3\vf m^2\omega_m^2}\int^\infty_0 d|\bQ| |\bQ|^3\int^\infty_{-\infty} 
d\Omega U^{\text{dyn}}(\bQ,\Omega_l)\left[\Omega_l
\tan^{-1}\frac{\vf |\bQ|}{\Omega_l}-(\Omega_l+\omega_m)\tan^{-1}
\frac{\vf|\bQ|}{\Omega_l+\omega_m}\right],\nn\\
\label{sc_3d}
\eea
whereas the dynamic interaction is approximated by 
\bea
U^{\text{dyn}}(\bQ,\Omega_l)=\frac{\pi}{2}N_F^{-1}\frac{\kappa^4}{(\bQ^2+\kappa^2)^2} \frac{|\Omega_l|}{\vf|\bQ|}.
\eea
The rest of the calculations is similar to the 2D case, except for the integral over $|\bQ|$ is not logarithmic. After simple algebra, we find
\bea
\delta\chi_{sc}(\omega_m)=-\frac{1}{24\pi^2}\frac{e^4}{\vf^2} k_{\text{F}} |\omega_m|.
\eea
Upon analytic continuation, this gives the bottom line in Eq.~(\ref{sc_low}) of the main text.
\subsubsection{Intermediate frequencies: $\vf\kappa\ll\omega\ll E_{\text{F}} $}
\label{app:sc_interm}
\paragraph{2D.}
Extension to frequencies in the intermediate region of $\vf\kappa\ll\omega\ll E_{\text{F}} $ is done in the same way as in Sec.~\ref{app:s_interm}: one only has to replace $\kappa$ in the lower limit of the logarithmic integral over $Q$  by $|\omega_m|/\vf$. Then the result can be written as
\bea
\delta\chi_{sc}(\omega_m)=-\frac{1}{6\pi^2}\frac{e^4}{\vf^2} |\omega_m|\ln\frac{E_{\text{F}} }{|\omega_m|}.
\eea
Upon analytic continuation, this gives Eq.~(\ref{sc2di}) of the main text.

\paragraph{3D.} As it was the case with the spin susceptibility (cf. Sec.~\ref{app:s_interm}), extension to the intermediate range of frequencies in 3D requires more work. As before, we approximate the dynamic interaction in Eq.~(\ref{sc_3d}) by Eq.~(\ref{Udi}). Folding the integral over $\Omega$ from $(-\infty,\infty)$ to $(0,\infty)$, we arrive at
\bea
\delta\chi_{sc}(\omega_m)=\frac{\kappa^4}{2\pi^3\vf m^2}\int^{E_{\text{F}} /\omega_m}\frac{dy}{y^2}f(y),
\eea
where $f(y)$ is given by Eq.~(\ref{A33}). Using the asymptotic expansion of $f(y)$ for large $y$ given by Eq.~(\ref{A35}), we find
\bea
\delta\chi_{sc}(\omega_m)=\frac{\ln 2}{2\pi^2}\frac{\kappa^4}{\vf m^2}\ln\frac{E_{\text{F}} }{|\omega_m |}.
\eea
We see that $\im\delta\chi_{sc}(\omega)$ is independent of $\omega$ in this frequency interval and given by Eq.~(\ref{sc3di}) of the main text.
\subsubsection{High-frequency region: $\omega\gg E_{\text{F}} $}
\label{app:sc_high}
At the first step, we integrate the product of the Green's functions in Eq.~(\ref{sc}) over $\e_m$ and shift the fermionic momenta in such way that all the Fermi functions are reduced to $n_F(\epsilon_\bk)$:
\bea
\chi_{sc}(\omega_m)&=&\frac{1}{m^2}\int\int\int\frac{d^2Qd\Omega_ld^2k}{(2\pi)^{
5
}}n_F(\epsilon_\bk) \bQ^2U^{\text{dyn}}(\bQ,\Omega_l)\nn\\
&&\times\left\{
\frac{1}{\left[i(\Omega_l+\omega_m)-\epsilon_{\bk+\bQ}+\epsilon_\bk\right]}\frac{1}{\left[i\Omega_l-\epsilon_{\bk+\bQ}+\epsilon_\bk\right]^2}
-\frac{1}{\left[i(\Omega_l+\omega_m)-\epsilon_{\bk}+\epsilon_{\bk-\bQ}\right]}\frac{1}{\left[i\Omega_l-\epsilon_{\bk}+\epsilon_{\bk-\bQ}\right]^2}
\right\}.\nn\\
\eea
As in Sec.~\ref{app:spin_high}, we first assume and then verify that $\bQ^2/m\gtrsim \Omega_l \sim \omega_m$. With this assumption, the differences of the dispersions in the equation above can be replaced by  $\epsilon_{\bk\pm\bQ}-\epsilon_\bk\approx \bQ^2/2m$. After this step, the integral over $\bk$ trivially gives the total number density, $n_0$. Then we have
\bea
\chi_{sc}(\omega_m)&=&\frac{n_0}{2m^2}\int\int\frac{d^2Qd\Omega_l}{(2\pi)^{
3
}} \bQ^2U^{\text{dyn}}(\bQ,\Omega_l)\nn\\
&&\times\left\{
\frac{1}{\left[i(\Omega_l+\omega_m)-\frac{\bQ^2}{2m}\right]}\frac{1}{\left[i\Omega_l-\frac{\bQ^2}{2m}\right]^2}
-\frac{1}{\left[i(\Omega_l+\omega_m)+\frac{\bQ^2}{2m}\right]}\frac{1}{\left[i\Omega_l+\frac{\bQ^2}{2m}\right]^2}
\right\}.\nn\\
\label{sc2}
\eea
Substituting Eq.~(\ref{Ud2}) for the dynamic interaction into Eq.~(\ref{sc2}) and rescaling the variables as 
$x=\Omega_l/\omega_m$ and $y=\bQ^2/2m\omega_m$ (with $\omega_m>0$), we obtain \bea
\delta\chi_{sc}=\frac{n_0^2e^4}{m\omega_m^2} \int^\infty_{E_{\text{F}} /\omega_m}\frac{dy}{y}
\int^\infty_{-\infty} dx\frac{x^2}{x^2+y^2}\left\{\frac{1}{\left[i(x+1)-y\right](ix-y)^2}-\frac{1}{\left[i(x+1)+y\right](ix+y)^2}\right\}.
\eea
As before, we kept the lower limit of the integral over $|\bQ|$ at $|\bQ|\sim k_{\text{F}} $ in anticipation of a logarithmic divergence. The integral over $x$ is equal to $\pi/(4y^2+1)$, and we indeed arrive at a logarithmic integral over $x$
\bea
\delta\chi_{sc}(\omega_m)=\frac{\pi n_0^2e^4}{m\omega_m^2} \int^\infty_{E_{\text{F}} /\omega_m}\frac{dy}{y}\frac{1}{4y^2+1}\approx \frac{\pi n_0^2e^4}{m\omega_m^2} \ln\frac{\omega_m}{E_{\text{F}} }.
\eea
As $\delta\chi_{sc}(\omega_m)$ must be an even function of $\omega_m$, it is obvious that for an arbitrary sign of $\omega_m$ the result reads
\bea
\delta\chi_{sc}(\omega_m)=\frac{\pi n_0^2e^4}{m\omega_m^2} \ln\frac{|\omega_m|}{E_{\text{F}} }.
\eea
Performing analytic continuation, expressing $n_0$ in terms of the Fermi energy via $n_0=mE_{\text{F}} /\pi$,  and taking the imaginary part, we arrive at Eq.~(\ref{sc_high}) of the main text.

A similar analysis shows that in 3D $\im\delta\chi_{sc}(\omega)\propto \text{sgn}\omega/|\omega|^{5/2}$.

\subsection{Contribution of the Aslamazov-Larkin diagrams to the charge susceptibility}
\label{sec:AL}
As we mentioned in the main text, the Aslamazov-Larkin (AL) diagrams in the spin channel vanish identically due to spin traces. However, they give a non-zero contribution to the irreducible charge susceptibility.
In this section, we show that in a Galilean-invariant system the AL diagrams cancel the leading contributions from the self-energy and ladder diagrams for $\omega\gg \vf |\bq|$. 

The sum of two AL diagrams (Fig.~\ref{fig1}, 4 and 5) can be written as
\begin{align}
\label{A60}
\delta\chi_{\mathrm{AL}}(\bq,\omega_m)=\chi^{(4)}(\bq,\omega_m)+\chi^{(5)}(\bq,\omega_m)=4\int_{{\bf Q},\Omega_l}\left[{\cal T}^2({\bf Q},{\bf q},\Omega_l,\omega_m)
+ \vert {\cal T}({\bf Q},{\bf q},\Omega_l,\omega_m)\vert^2\right]
U({\bf Q}-{\bf q},\Omega_l-\omega_m)U({\bf Q},\Omega_l),
\end{align}
where a factor of $4$ is due to the trace over spins, and
\begin{align}
{\cal T}({\bf Q},{\bf q},\Omega_l,\omega_m) = \int_{\bk,\e_n}G({\bf k},\e_n) G({\bf k} + {\bf q},\e_n+\omega_m)
G({\bf k}+{\bf Q},\e_n + \Omega_l)\label{T}
\end{align}
is a ``triangle'' formed by three Green's functions.

First, we prove that $\delta\chi_{\mathrm{AL}}(\bq\to {\bf 0},\omega_m)=0$. This condition guarantees charge conservation as we already know that the sum of the self-energy and ladder diagrams does vanish at $\bq=0$. The two AL diagrams cancel each other  because \bea
{\cal T}^*({\bf Q},{\bf 0},\Omega_l,\omega_m)=-{\cal T}\label{Tim}({\bf Q},{\bf 0},\Omega_l,\omega_m)\eea and thus  ${\cal T}^2+|{\cal T}|^2=0$ at $\bq=0$. To see this in more detail, we put $\bq=0$ in Eq.~(\ref{T}), apply identity (\ref{a6}) to the first two Green's functions in ${\cal T}$, and recall that $\Pi({\bf Q},\zeta_m)=\int_{\bk,\e_n} G(\bk+\bQ,\e_n+\zeta_m)G(\bk,\e_n)$ is the polarization bubble on the Matsubara axis. This yields
\bea
{\cal T}({\bf Q},{\bf 0},\Omega_l,\omega_m)=\frac{1}{i\omega_m}\left[\Pi(\bQ,\Omega_l)-\Pi(\bQ,\Omega_l-\omega_m)\right].\label{T0}
\eea
We now recall that $\Pi({\bf Q},\zeta_m)$ is purely real and thus ${\cal T}({\bf Q},{\bf 0},\Omega_l,\omega_m)$ is purely imaginary, which proves our assertion, Eq.~(\ref{Tim}).
[That $\Pi(\bQ,\zeta_m)$ is purely real follows from the spectral representation $\Pi(\bQ,\zeta_m)=(1/\pi)\int dz \im \Pi(\bQ,z)/(z-i\zeta_m)$ and the condition that $\im \Pi(\bQ,z)$ is an odd function of $z$.]

The same requirement, i.e., that any susceptibility on the Matsubara axis must be  a real-valued quantity, implies that the imaginary part of ${\cal T}^2$ (at any $\bq$) must vanish on subsequent integrations, and thus ${\cal T}^2+\vert {\cal T}^2\vert$ must be reduced to $2\left( \mathrm{Re}{\cal T} \right)^2$. We will be using this observation later on.

Finally, we first assume and then verify that the Coulomb interaction can be approximated by its static form. Also, since we are free to choose the external momentum $\bq$ such that $|\bq|\ll\kappa$ while we expect that $|\bQ|\sim \kappa$, we can also neglect $\bq$ in one of the Coulomb potentials. With these simplifications,
\begin{align}
U({\bf Q}-{\bf q},\Omega-\omega)U({\bf Q},\Omega) \approx \left(\frac{2\pi e^2}{|\bQ| + \kappa}\right)^2.
\end{align}

Now we calculate the triangle ${\cal T}$. Integrating over $\e_n$, we obtain
\begin{align}
\label{A32}
{\cal T}({\bf Q},{\bf q},\Omega_l,\omega_m)= -
\int \frac{d^2k}{(2\pi)^2}
\frac{1}{\omega_m +
i\left(\epsilon_{\bk+\bQ} -\epsilon_\bk\right)
}
\left[ \frac{n_F\left(\epsilon_{\bf k}\right) - n_F\left(-\epsilon_{{\bf k} + {\bf Q}}\right)}{\Omega_l + i\left(\epsilon_{\bk+\bQ} - \epsilon_\bk\right)}
 - \frac{n_F\left(-\epsilon_{\bk+\bq}\right) - n_F\left(-\epsilon_{\bk+\bQ}\right)}{\Omega_l - \omega_m + i\left(\epsilon_{\bk+\bQ} - \epsilon_{\bk+\bq} \right)}\right].
 \end{align}
At this point, the calculation deviates from the procedure which we used to calculate the self-energy and ladder diagrams. Namely, if we expand dispersions to linear order in $\bq$ and $\bQ$, as we did in previous two sections, the entire ${\cal T}$ would be purely imaginary and thus the ${\cal T}^2$ and $|{\cal T}|^2$ terms would cancel each other not only at $\bq=0$ but also at finite $\bq$. To get a non-zero result, we need to keep $\bQ^2/2m$ terms in the dispersions. However, $\bq^2/2m$ terms can still be neglected because $\bq$ can be chosen arbitrarily small.

Shifting the momenta in Eq.~(\ref{A32}) in such a way that all the Fermi functions become $n_F\left(\epsilon_\bk\right)$ and neglecting ${\cal O}(\bq^2)$ terms, we get
\begin{align}
&{\cal T}({\bf Q},{\bf q},\Omega_l,\omega_m) = -\int^{\kf}_0 \frac{d|\bk||\bk|}{2\pi} \int \frac{d\theta_{\bk{\bf Q}}}{2\pi}
\bigg[ \frac{1}{\omega_q\Omega_+}
-\frac{1}{\left(\omega_q- \frac{i}{m}{\bf Q}\cdot{\bf q} \right)\Omega_-}
\nn\\
&
-\frac{1}{\omega_q\left( \Omega_+ - \omega_q + \frac{i}{m}{\bf Q}\cdot{\bf q}\right)}
+ \frac{1}{\left(\omega_q - \frac{i}{m}{\bf Q}\cdot{\bf q} \right) \left( \Omega_ - -\omega_q + \frac{i}{m}{\bf Q}\cdot{\bf q}\right)}\bigg],\label{A33}
\end{align}
where
\begin{align}
&\omega_{q}= \omega_m +i\vf\hat\bk\cdot\bq,\nn\\
&\Omega_{\pm} = \Omega_l \pm i\left(\epsilon_{\bk \pm {\bf Q}} - \epsilon_\bk\right).
\end{align}
Since  $|\omega|\gg \vf |\bq|$ by assumption, we can replace $\omega_{q}$ by $\omega_m$  
everywhere. Next, we expand the integrand in Eq.~(\ref{A33}) to first order in $\frac{1}{m}{\bf Q}\cdot{\bf q}$, which yields
\begin{align}
&{\cal T}({\bf Q},{\bf q},\Omega_l,\omega_m)=
{\cal T}({\bf Q},{\bf 0},\Omega_l,\omega_m)
 - \frac{i}{m}{\bf Q}\cdot{\bf q}\int^{k_{\text{F}} }_0 \frac{dkk}{2\pi} 
\int \frac{d\theta_{\bk{\bf Q}}}{2\pi}\bigg[ 
 \frac{1}{\omega_m^2}\left(\frac{1}{\Omega_{-} - \omega_m} - \frac{1}{\Omega_{-} }  \right)
- 
\frac{1}{\omega_m}\left(\frac{1}{(\Omega_+ - \omega_m)^2}
+ \frac{1}{(\Omega_{-} - \omega_m)^2}\right)  \bigg],
\label{tp}
\end{align}
where the (purely imaginary) leading term ${\cal T}({\bf Q},{\bf 0},\Omega_l,\omega_m)$ is given by Eq.~(\ref{T0}).

Now we can see why linearization of the electron spectrum gives a zero result for the AL diagrams. Suppose that we perform such linearization upon which $\Omega_+=\Omega_-=\Omega_l+i\vf\hat\bk\cdot\bQ$. The angular integrals  in Eq.~(\ref{tp}) are calculated using
\bea
I(z)&=&\int^{2\pi}_0\frac{d\phi}{2\pi}\frac{1}{iz-\cos\phi}=-i\frac{\text{sgn}z}{\sqrt{z^2+1}},\nn\\
J(z)&=&\int^{2\pi}_0\frac{d\phi}{2\pi}\frac{1}{\left(iz-\cos\phi\right)^2}=i\frac{\partial}{\partial z}I(z)=
-\frac{|z|}{\left(z^2+1\right)^{3/2}},\label{ints}
\eea
where $z$ is purely real.
(In the second line, we neglected the $\delta(z)$ term which gives no contribution to the imaginary part of the susceptibility.)
It is easy to see that the contributions to ${\cal T}$ from both the first and second terms in the square brackets in Eq.~(\ref{tp}) are purely imaginary  in this approximation, and thus the two AL diagrams cancel each other.

To get a non-zero result, we need to keep the $\bQ^2/2m$ terms in the dispersions, i.e.,
take $\Omega_{\pm}$ as 
\bea
\Omega_{\pm}=\Omega_l+i\left(\frac{1}{m}\bk\cdot\bQ\pm \frac{\bQ^2}{2m}\right).
\eea
Keeping the $\bQ^2/2m$ term amounts to shifting $z$ in Eq.~(\ref{ints}) by a complex number, which gives a real-valued correction to the integral in the top line and imaginary-valued correction to the integral in the bottom one. However, such a correction will cancel out between the two terms in the second round bracket in the integrand of Eq.~(\ref{tp}). Therefore, we need to keep only the first round bracket in there.
The rest of integrations is performed as follows:
\begin{align}
&{\cal T}({\bf Q},{\bf q},\Omega_l,\omega_m)-{\cal T}({\bf Q},{\bf 0},\Omega_l,\omega_m)= -\frac{i{\bf Q}\cdot{\bf q}}{m\omega_m^2}
\int^{k_{\text{F}} }_0 \frac{d|\bk||\bk|}{2\pi} \int \frac{d\theta_{\bk{\bf Q}}}{2\pi}
\left(\frac{1}{\Omega_{-} - \omega_m} - \frac{1}{ \Omega_{+} }  \right)\nn
\\
&
= -\frac{i{\bf Q}\cdot{\bf q}}{m\omega_m^2} \int_{0}^{k_{F}} \frac{d|\bk||\bk|}{2\pi}
\left[ \frac{\text{sgn}(\Omega - \omega)}{\sqrt{ (\Omega - \omega - i\frac{\bQ^2}{2m})^2 + \left( \frac{|\bk|}{m}\right)^2\bQ^2}}
-\frac{\text{sgn}(\Omega)}{\sqrt{ (\Omega - i\frac{Q^2}{2m})^2 + \left( \frac{|\bk|}{m}\right)^2\bQ^2}}  \right]\nn
\\
&
= -\frac{im{\bf Q}\cdot{\bf q}}{2\pi\omega_m^2\bQ^2}
\left\{
\mathrm{sgn}(\Omega_l - \omega_m) \left[ \sqrt{ \left(\Omega_l - \omega_m - i\frac{\bQ^2}{2m}\right)^2 + \vf^2\bQ^2} - \sqrt{ \left(\Omega_l - \omega_m - i\frac{\bQ^2}{2m}\right)^2 } \right]\right.\nn
\\
&
\left. -
\mathrm{sgn}(\Omega_l) \left[ \sqrt{ \left(\Omega_l  - i\frac{\bQ^2}{2m}\right)^2 + \vf^2\bQ^2} - \sqrt{ \left(\Omega_l  - i\frac{\bQ^2}{2m}\right)^2 } \right]\right\}.
\end{align}
Now, we may safely expand the expression above to first order in $i\bQ^2/2m$; this gives a real-valued correction to ${\cal T}$. When ${\cal T}$ is squared, the cross-term of ${\cal T}({\bf Q},{\bf 0},\Omega_l,\omega_m)$   and the ${\cal O}(\bQ\cdot\bq)$ correction will vanish on integrating over the angle between $\bq$ and $\bQ$ (such a term is purely imaginary and thus must vanish anyway). Keeping only that part of ${\cal T}$ which gives an ${\cal O}(\bq^2)$ contribution to $\im\chi_{\text{AL}}(\bq,\omega_m)$, we obtain
\begin{align}
{\cal T}= -  \frac{{\bf Q}\cdot{\bf q}}{4\pi \omega^2}
\left[ \frac{\vert \Omega_l - \omega_m\vert}{\sqrt{(\Omega_l - \omega_m)^2 +\vf^2\bQ^2 }}
- \frac{\vert \Omega_l \vert}{\sqrt{\Omega_l^2 + \vf^2\bQ^2 }} \right].
\end{align}
Substituting this result back into Eq~.(\ref{A60}), and rescaling the variables as $x=\Omega_l/\vf|\bQ|$ and $y=\omega_m/\vf|\bQ|$, we find
\begin{align}
 &\delta\chi
_{\mathrm{AL}}(\bq,\omega_m)= \frac{\vf}{2\pi^2\omega_m^4}\int \frac{d^2Q}{(2\pi)^2}
%
\frac{(2\pi e^2)^2}{(|\bQ|+\kappa)^2} ({\bf Q}\cdot{\bf q})^2|\bQ|F\left(\frac{\omega_m}{\vf|\bQ|}\right),
\end{align}
where
\bea
F(y)=\int^{\infty}_{-\infty} dx \left(\frac{|x-y|}{\sqrt{(x-y)^2+1}}-\frac{|x|}{\sqrt{x^2+1}}\right)^2.\eea
We are interested in the limit of $y\ll 1$, when
\bea
F(y)=
\frac{3\pi}{8}y^2-\frac 23 |y|^3+{\cal O}(y^4).\label{A74}
\eea
Upon analytic continuation, the first term in Eq.~(\ref{A74}) contributes only to the real part of the susceptibility and thus will be discarded. Keeping only the second term and integrating over $\bQ$, we obtain the final result
for the AL contribution \begin{align}
&\delta\chi
_{\mathrm{AL}}(\bq,\omega_m) =-\lambda_2 \frac{\bq^2}{|\omega_m|},
\end{align}
where $\lambda_2$ is given by Eq.~(\ref{lambda2}). 
Upon analytic continuation, the last result gives Eq.~(\ref{tail}) of the main text.


\end{widetext}
\bibliography{dm_references}
\end{document}